\documentclass[a4paper, allowtoday, 11pt]{quantumarticle}
\pdfoutput=1
\usepackage[dvipsnames, svgnames]{xcolor}
\usepackage[utf8]{inputenc}
\usepackage{hyperref}
\usepackage{changes}
\hypersetup{
    colorlinks=true}
\usepackage[british]{babel}
\usepackage[inline]{enumitem}
\usepackage[normalem]{ulem}
\usepackage{physics}
\usepackage{verbatim}
\usepackage[operators,sets]{cryptocode}
\usepackage{bbm}
\usepackage{amsmath}
\usepackage{amssymb}
\usepackage{mathtools}
\usepackage{mathrsfs, dsfont} 
\usepackage{multirow}
\usepackage{tikz}
\usepackage{tikz-3dplot}
\usepackage{tabularray}
\usepackage{booktabs}
\usetikzlibrary{decorations.pathreplacing,calligraphy, 3d, backgrounds, fit, positioning, quantikz2}
\usepackage{graphicx}
\usepackage{subcaption}
\usepackage{ragged2e}
\DeclareCaptionJustification{justified}{\justifying}

\usepackage{amsthm}

\newtheorem{definition}{Definition}

\usepackage{algorithm}
\usepackage{algorithmic}
\usepackage{enumitem}

\floatstyle{ruled}

\newcommand{\Id}{\mathsf{I}}
\newcommand{\X}{\mathsf{X}}
\newcommand{\Y}{\mathsf{Y}}
\newcommand{\Z}{\mathsf{Z}}
\renewcommand{\S}{\mathsf{S}}

\newcommand{\Ha}{\mathsf{H}}
\newcommand{\CZ}{\mathsf{CZ}}

\newcommand{\MZZ}{\mathsf{M}_{\Z\Z}}
\newcommand{\MXX}{\mathsf{M}_{\X\X}}
\newcommand{\MX}{\mathsf{M}_{\X}}
\newcommand{\MYY}{\mathsf{M}_{\Y\Y}}

\newcommand{\circuit}{\ensuremath{\mathscr{C}}}

\newcommand{\bunch}{\texttt{Bunch}}

\newcommand{\suc}{\texttt{Success}}
\newcommand{\abo}{\texttt{Abort}}
\newcommand{\sucp}{$\mathtt{Success_+}$}
\newcommand{\sucm}{$\mathtt{Success_-}$}
\newcommand{\loss}{\texttt{Loss}}
\newcommand{\lossone}{$\mathtt{Loss_1}$}
\newcommand{\losstwo}{$\mathtt{Loss_2}$}

\newcommand{\fspoqc}{$\MZZ$-SPOQC}

\definecolor{hcgreen}{RGB}{101,163,66}
\definecolor{hcblue}{RGB}{48,111,183}
\definecolor{hcred}{RGB}{234,51,35}

\definecolor{quandela_blue}{HTML}{7AA0D2}
\definecolor{quandela_light_blue}{HTML}{7AB5DB}
\definecolor{quandela_purple1}{HTML}{8888C6}
\definecolor{quandela_purple2}{HTML}{946CBA}

\tikzset{
    3Dstack/.style={
        x={(1cm, 0cm)},
        y={(0cm, 1.2cm)},
        z={(3mm, 5mm)},
    },
    my triangle/.style={-{Triangle[width=\the\dimexpr1.8\pgflinewidth,length=\the\dimexpr1.5\pgflinewidth]}},
    B/.style = {decorate,
            decoration={calligraphic brace, amplitude=4pt,
            raise=1pt, mirror},
            very thick,
            pen colour=black},
}

\allowdisplaybreaks

\begin{document}

\title{Enhanced Fault-tolerance in Photonic Quantum Computing:
Comparing the Honeycomb Floquet Code and the Surface Code in Tailored Architecture
}
\author{Théo Dessertaine}
\affiliation{Quandela, 7 Rue Léonard de Vinci, 91300 Massy, France}

\author{Boris Bourdoncle}
\affiliation{Quandela, 7 Rue Léonard de Vinci, 91300 Massy, France}

\author{Aurélie Denys}
\affiliation{Quandela, 7 Rue Léonard de Vinci, 91300 Massy, France}

\author{Grégoire de Gliniasty}
\affiliation{Quandela, 7 Rue Léonard de Vinci, 91300 Massy, France}
\affiliation{Sorbonne Université, CNRS, LIP6, F-75005 Paris, France}

\author{Pierre Colonna d'Istria}
\affiliation{Quandela, 7 Rue Léonard de Vinci, 91300 Massy, France}

\author{Gerard Valent\'i-Rojas}
\affiliation{Naquidis Center, Institut d’Optique Graduate School, 91127, Palaiseau, France}
\affiliation{Institut de Mathématiques de Bordeaux, UMR 5251, Université de Bordeaux, France}

\author{Shane Mansfield}
\affiliation{Quandela, 7 Rue Léonard de Vinci, 91300 Massy, France}

\author{Paul Hilaire}
\affiliation{Quandela, 7 Rue Léonard de Vinci, 91300 Massy, France}

\begin{abstract}
Fault-tolerant quantum computing is crucial for realizing large-scale quantum computation, and the interplay between hardware architecture and quantum error-correcting codes is a key consideration. We present a comparative study of two quantum error-correcting codes -- the surface code and the honeycomb Floquet code -- implemented on the spin-optical quantum computing architecture, either with controlled-Z operations or with direct parity measurements. This allows for a direct comparison of the codes using consistent noise models. Notably, we achieve a loss threshold of 6.3\% with the honeycomb Floquet code implemented on our tailored architecture, almost twice as high as the loss threshold obtained with the surface code on the previous architecture, all the while requiring less physical qubits. This finding is particularly significant given that photon loss is the primary source of errors in photon-mediated quantum computing. Moreover, we benchmark the general performances of the two codes in a multi-error setting by computing the volume of the fault-tolerant region, and show that the fault-tolerant region of the honeycomb code is over twice as large as that of the surface code.

\end{abstract}

\maketitle
\section{Introduction}

Fault-tolerant quantum computing aims to reliably process quantum information, even in the presence of physical noise, thanks to the threshold theorem \cite{threshold_theorem}. Fault tolerance via quantum error correction is necessary to achieve large-scale quantum computation capable of implementing the kinds of high-impact applications that would derive for instance from Shor's factoring algorithm \cite{shor}, or the Harrow-Hassidim-Lloyd algorithm \cite{hhl} for solving linear systems. In the pursuit of practical fault-tolerant quantum computing, the interplay between hardware architecture and quantum error-correcting codes is crucial.
In this article we evidence the point with a comparative performance study of two kinds of code – the surface code \cite{bravyi1998quantum} and the honeycomb Floquet code \cite{HH2021Dynamically} – when implemented with two variants of the spin-optical quantum computing (SPOQC) architecture \cite{Gliniasty2024Spin}. 

This modular architecture has been designed specifically for computing platforms consisting of quantum-dot photon emitters~\cite{lee2019quantum,  appel2022entangling, coste2023high,  cogan2023deterministic, meng2023photonic, su2024continuous}, photon routers and small linear optical circuits. These same basic modules have already been used to realize small-scale noisy quantum computing platforms \cite{maring2024versatile}. However, the architecture may equally be realized using other kinds of photon emitters that have an embedded qubit degree of freedom (e.g.~\cite{blinov2004observation, thomas2022efficient, OSullivan2024}). In the original version of the SPOQC architecture linear optical circuits were used to perform ``repeat-until-success'' (RUS) controlled-$\Z$ gates. These are especially convenient for running  error correction, for example via the popular surface code~\cite{bravyi1998quantum}. The surface code allows fault-tolerance~\cite{PhysRevA.80.052312} and is attractive for its versatility and simplicity, admitting a 2D planar layout and good error thresholds. In particular it has recently been implemented at small scale below threshold on real hardware~\cite{acharya2024quantum}. Implemented on the SPOQC architecture, the surface code~\cite{Gliniasty2024Spin} was shown to allow a photon loss threshold of $2.8\%$ \footnote{Here, we improve on this value thanks to a better understanding of the noise model induced by photon loss.}.

However, a limitation of the surface code is that it relies on the measurement of weight-$4$ Pauli operators to check for errors, and high-weight measurements are challenging to implement in practice. A typical way to circumvent this consists in using syndrome extraction circuits, in which four two-qubit entangling gates and an auxiliary qubit allow for indirect measurement of the check operator~\cite{devitt2013quantum}.

In that regard, the recent development of a new family of codes known as Floquet codes~\cite{alam2024dynamical, KMT2024Anyon, HB2023Constructions, FDB2024Fault, TMK2023Floquetifying, DTB2023Floquet, ustun2024single} opens exciting new possibilities.
Through periodic sequences of weight-$2$ measurements, such codes dynamically generate effective stabilizers of higher weight over time.
They can thus achieve the benefits of higher-weight check operators while only requiring simpler lower-weight measurements. This makes them well-suited to technological platforms for which such operations are native, i.e.\ easy to implement, such as photonic platforms~\cite{paesani2023high} or Majorana-based qubits~\cite{PKD2023Performance}.
Of particular interest is the honeycomb Floquet code ~\cite{HH2021Dynamically}, which may be understood as a dynamical version, or ``floquetification'', of the surface code, in the sense that it switches between three implementations of the surface code. A previous comparison indicates that the honeycomb code  performs significantly better than the surface code on hardware with native two-qubit measurements~\cite{GNFB2021Fault}.
However, the comparison relied on different noise models for the native operations used to implement the respective codes.

In this article, we propose a variant on the original SPOQC architecture \cite{Gliniasty2024Spin} tailored to two-qubit measurements. Crucially, the same physical noise model applies to the original architecture and the present variant, respectively referred to as $\CZ$-SPOQC and $\MZZ$-SPOQC in the following. This allows for a fair comparison of the performance of quantum error-correcting (QEC) codes on these variants. Remarkably, we observe a photon loss threshold of $6.3\%$ with the honeycomb code implemented of the present tailored architecture, almost twice as large as the previous threshold obtained for the surface code, and we achieve so without relying on large-scale multiplexing~\cite{bartolucci2023fusion, bartolucci2021creation, omkar2022all, bombin2023increasing, pankovich2024flexible, pankovich2024high, lobl2024loss}. This increase in loss tolerance is of particular importance as photon loss is the dominant source of errors in photon-mediated quantum computing platforms, and multiplexing strategies significantly increase the resource overhead of photonic fault-tolerant architectures. Moreover, we show that the surface code, when implemented on the $\CZ$-SPOQC variant, is outperformed by the honeycomb code implemented on the $\MZZ$-SPOQC variant, thus demonstrating the importance of co-design between QEC codes and their physical implementations. Finally, we analyze an implementation of the surface code on the $\MZZ$-SPOQC variant,  which exhibits a surprising resilience against photon distinguishability errors.

\paragraph*{\textnormal{\textbf{Related work.}}}
Several FTQC architectures based on quantum emitters have been proposed recently. Notable examples include~\cite{Gliniasty2024Spin}, which leverages RUS gates between quantum emitters to implement CZ gates, and~\cite{paesani2023high}, which introduces a Floquet color code approach in the fusion-based quantum computing (FBQC) framework~\cite{bartolucci2023fusion}, using linear cluster states generated by the quantum emitters. While preparing this manuscript, we became aware of a related but independent work on fault-tolerant spin-optical architectures using dynamical codes and parity measurements~\cite{chan2024tailoring}. 

The work of Chan \textit{et al.} \cite{chan2024tailoring} focuses on the CSS honeycomb code introduced in \cite{KMT2024Anyon, DTB2023Floquet} and compares two schemes for implementing the parity measurements, one based on the RUS mechanism and one using boosted fusion gates~\cite{hilaire2023near}. The present work investigates two QEC codes, the surface code~\cite{bravyi1998quantum, Gidney2023Pair} and the Hastings-Haah honeycomb code with boundary conditions~\cite{HH2022Boundaries,GNM2022Benchmarking}, together with two types of entangling operations, namely CZ gates and parity measurements. For the four combinations of code and entangling operation, we compute the volume of the fault-tolerant region in order to compare their performance. The two articles also differ in their simulation methods. Both investigate the same three sources of noise that are relevant to quantum-emitter-based architectures: photon loss, spin decoherence, and photon distinguishability. However, Chan \textit{et al.} construct a custom framework to propagate emitter-level noise into measurement errors and build the decoding graph directly from the noise model, whereas the present work performs full circuit simulations using the Clifford circuit simulator Stim~\cite{gidney2021stim}.

\section{Two variants of a spin-optical architecture for FTQC} 

Implementing a quantum error-correcting code requires preparing the logical states of the code and performing syndrome extraction. For the very large class of stabilizer codes, this can be achieved using only controlled-$\Z$ ($\CZ$) gates, the change-of-basis gates $\Ha= (\X+\Z)/\sqrt{2}$ and $\Ha_{\Y\Z}=(\Y+\Z)/\sqrt{2}$, along with initialization and measurement in the computational basis. Data qubits that carry the logical information are entangled with auxiliary qubits whose measurement is equivalent to measuring a given stabilizer operator. This simple scheme is widely used but can be resource intensive, as it requires additional qubits as well as several entanglement layers. Dynamical codes, however, are more naturally implemented through direct two-qubit measurements. In this setting, there is \emph{a priori} no need for additional qubits, since we directly perform entangling two-qubit measurements on the data qubits. 

In both cases, one needs to define the physical operations that implement the required gates and measurements on the qubits of our system. The spin-optical architecture SPOQC introduced in \cite{Gliniasty2024Spin} enables to perform $\CZ$ gates using a repeat-until-success (RUS) subroutine, see below. In this hybrid architecture, quantum information is encoded on the spin degree of freedom of quantum emitters. Upon emission, these spins are entangled with photonic qubits and some gates can be indirectly applied on the spin qubits by means of operations performed on the photonic qubits, including photon detections. Here, we introduce a variant of SPOQC, which directly implements RUS two-qubit $\Z\Z$ measurements, denoted RUS $\MZZ$, making it suitable for dynamical codes. 

\subsection{The RUS subroutine}
\label{sec:RUS}

\begin{figure}
    \centering
    \includegraphics[width=\linewidth]{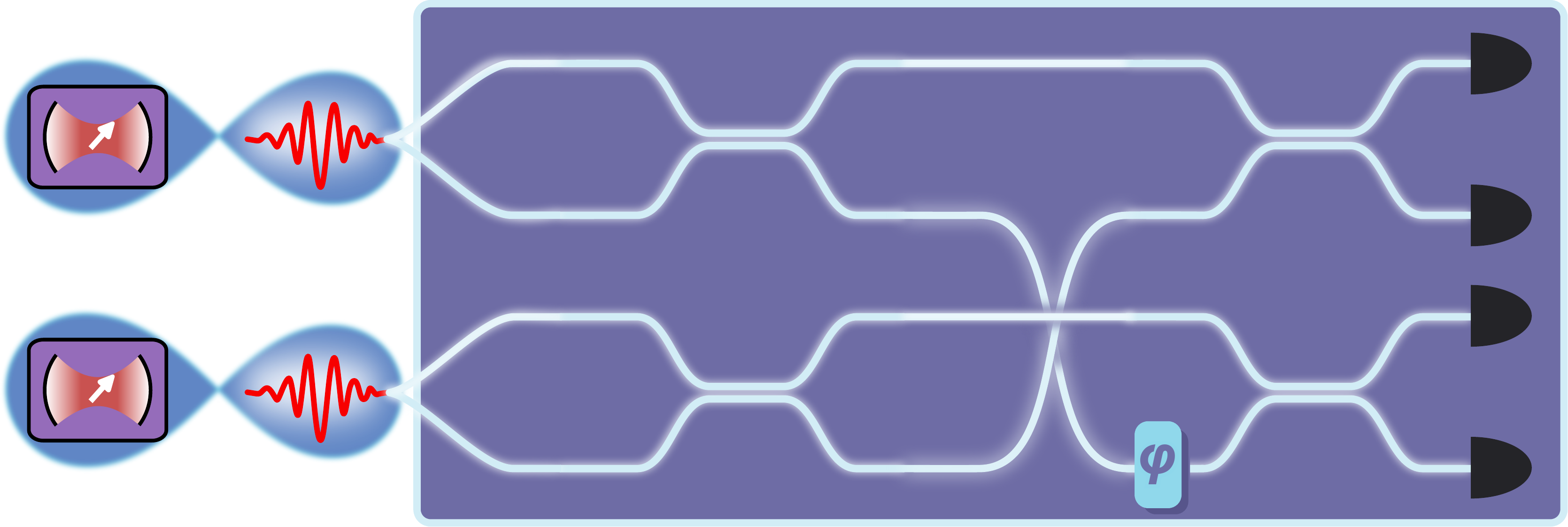}
    \caption{Linear-optical interferometer implementing the unitary $U(\varphi)$ used in the RUS subroutine. 
    Optical mode brought together implement balanced beam-splitters and the light-blue rectangle is a phase-shifter adding a phase $\varphi$.
    The nature of the operation depends on $\varphi$: if $\varphi=\pi/2$, the target operation is a $\CZ$ gate; if $\varphi=0$, it's an $\MZZ$ measurement of the spins.}
    \label{fig:interferometer}
\end{figure}

The subroutine for the RUS operation is as follows.
\begin{enumerate}[label=(\arabic*), leftmargin=20pt]
    \item Each quantum emitter generates a spin-entangled photon through the process $\mathsf{E}$~\cite{lee2019quantum, thomas2022efficient, appel2022entangling, coste2023high,  cogan2023deterministic, meng2023photonic, su2024continuous, OSullivan2024}:
    \begin{equation}
        \mathsf{E} = \ket{0_{\rm s}, 0_{\rm ph}} \bra{0_{\rm s}} + \ket{1_{\rm s}, 1_{\rm ph}} \bra{1_{\rm s}}, \label{eq:emission process}
    \end{equation}
    where the subscripts ``s'' and ``ph'' respectively denote the spin qubit and dual-rail-encoded photonic qubit (see Appendix \ref{app:interferometer}). 
   \item The two spin-entangled photons  enter a $4$-mode linear-optical interferometer that performs a unitary transformation on its input optical modes.
    \item The photons are detected by photon-number-resolving detectors located at the end of each linear-optical interferometer output modes. Depending on the detection patterns, the spin qubits are projected onto different states and we may  perform single-qubit gate corrections on the spin qubits.
\end{enumerate}

In this work, we use the interferometer depicted in Fig.~\ref{fig:interferometer}. It implements a transformation $U(\varphi)$ on the input photonic modes that reads:
\begin{align}
U(\varphi)&=
\frac{1}{2}\begin{pmatrix}
    1 & 1 & 1 & -1 \\
    1 & 1 & -1 & 1 \\
    e^{i\varphi} & -e^{i\varphi} & 1 & 1 \\
    -e^{i\varphi} & e^{i\varphi} & 1 & 1
\end{pmatrix}.
\end{align}
When $\varphi=\pi/2$, it corresponds to a slight variant from the original interferometer in \cite{Gliniasty2024Spin}. The classical signal describing where photons were detected is called a measurement pattern. The detection applies a quantum channel on the spin qubits whose form depends on the pattern. In the ideal case of indistinguishable photons that don't suffer loss, the measurement patterns can be divided into two categories. The \bunch\ patterns correspond to the detection of two photons in the same output mode. In the case of $U(\varphi)$, the quantum channel associated to these patterns is simply the identity map $\mathcal{I}:\rho\mapsto\rho$, where $\rho$ denotes the density matrix of the two spin qubits. All of the channels are considered after single-qubit corrections see Appendix~\ref{sec:simu_rus_gate}. The \suc\ patterns correspond to the detection of valid dual-rail patterns, i.e. one photon is detected in one of the first two modes and one photon is detected in one of the last two modes. For these patterns, the quantum channel depends on the phase of the interferometer. More precisely, setting $\varphi=\pi/2$ implements a $\CZ$ operation, and thus corresponds to the $\CZ$-SPOQC architecture \cite{Gliniasty2024Spin}, while setting $\varphi=0$ implements the two-qubit $\Z\Z$ measurement needed for the $\MZZ$-SPOQC architecture (see Appendix \ref{app:interferometer} for details on the computation). The \suc\ and \bunch\ patterns have the same probability of occurring, and the subroutine is repeated until one of the \suc\ patterns is observed. In the non-ideal case, other patterns due to photon loss and distinguishability can be observed.

\subsection{Error model}
\label{sec:error-models}

\begin{figure*}
    \centering
    
    \begin{tikzpicture}

        \node[draw, very thick, circle, inner sep=2pt] (one) at ($(-5, 0)+(240:2.2)$) {$\pmb{(1)}$};
        \node[draw, very thick, circle, inner sep=2pt] (two) at ($(-5, 0)+(120:2.2)$) {$\pmb{(2)}$};
        \node[draw, very thick, circle, inner sep=2pt] (three) at ($(-5, 0)+(0:2.2)$) {$\pmb{(3)}$};
        \node[align=left] (a) at ($(two)+(120:1)+(0,0)$) {{\bf (a)} $\CZ$-SPOQC};
        
        \node[draw, fill=BrickRed!20, rounded corners=4pt] (fail) at ($(three)+(25:3)$) {\bf Lost phase};
        \node[BrickRed]  at ($(three)+(25:3)+(90:0.75)$) {$\mathcal{C}_{loss}^{(\infty)}$};
        \node[draw, fill=ForestGreen!20, rounded corners=4pt] (suc) at ($(three)+(-25:3)$) {\bf Noisy $\CZ$};
        \node[ForestGreen]  at ($(three)+(-25:3)+(-90:0.75)$) {$\mathcal{C}_{dist.}^{\CZ}$};
        \node[draw, fill=Gray!20, rounded corners=4pt] (abo) at ($(one)+(-90:1.8)$) {\bf No entanglement};

        \draw[-stealth, ultra thick] (one) to[bend left=45] (two);
        \draw[-stealth, ultra thick] (two) to[bend left=45] (three);
        \draw[-stealth, ultra thick, CornflowerBlue] (three) to[bend left=25] node[pos=0.6, label={[yshift=-14mm, align=center]:{ \bunch\\$\mathcal{I}$}}] {} (one);
        \draw[-stealth, ultra thick, Salmon] (three) to[bend left=-25] node[pos=0.4, label={[xshift=18mm, yshift=-17mm, align=center]:{\lossone\\$\mathcal{C}_{loss}^{(1)}$}}] {} (one);
        \draw[-stealth, ultra thick, Gray] (one) to node[pos=0.4, label={[xshift=-10mm, yshift=-5mm, align=center]:{\abo\\$\mathcal{C}_{abor.}^{\CZ}$}}] {} (abo);

        \draw[-stealth, ultra thick, BrickRed] (three) to node[midway, below right] {\small \losstwo} (fail);
        \draw[-stealth, ultra thick, ForestGreen] (three) to node[midway, above right] {\small \suc} (suc);

        \node[circle, fill=NavyBlue] (middle) at (-5,0) {\color{white}$T_2$};
        \node[NavyBlue] (middle) at ($(-5,0)+(90:0.75)$) {$\mathcal{C}_{deco.}^{(t_{\rm cycle})}$};
        \node[scale=2.7, rotate=-90, white] (middle_cycle) at (-5,0) {$\circlearrowright$};

        \node[draw, very thick, circle, inner sep=2pt] (one) at ($(4, 0)+(240:2.2)$) {$\pmb{(1)}$};
        \node[draw, very thick, circle, inner sep=2pt] (two) at ($(4, 0)+(120:2.2)$) {$\pmb{(2)}$};
        \node[draw, very thick, circle, inner sep=2pt] (three) at ($(4, 0)+(0:2.2)$) {$\pmb{(3)}$};
        
        \node[align=left] (b) at ($(two)+(120:1)$) {{\bf (b)} \fspoqc{}};

        \node[draw, fill=ForestGreen!20, rounded corners=4pt, align=center] (suc_p) at ($(three)+(0:2.5)$) {\bf Noisy \\$\mathsf{M}_{\Z\Z}$};


        \node[ForestGreen]  at ($(suc_p)+(90:0.85)$) {$\mathcal{C}_{dist.}^{\MZZ}$};
        
        \node[draw, fill=Gray!20, rounded corners=4pt] (abo) at ($(one)+(-90:1.8)$) {\bf No entanglement};

        \draw[-stealth, ultra thick] (one) to[bend left=45] (two);
        \draw[-stealth, ultra thick] (two) to[bend left=45] (three);
        \draw[-stealth, ultra thick, BrickRed] (three) to[bend left=25] node[BrickRed, pos=0.75, label={[yshift=-14mm, align=center]:{\losstwo\\$\mathcal{C}_{loss}^{(\infty)}$}}, align=center] {} (one);
        \draw[-stealth, ultra thick, CornflowerBlue] (three) to[bend left=0] node[CornflowerBlue, pos=0.5, label={[xshift=9mm, yshift=-15mm, align=center]:{\bunch\\$\mathcal{I}$}}] {} (one);
        \draw[-stealth, ultra thick, Salmon] (three) to[bend left=-25] node[Salmon, pos=0.25, label={[xshift=17mm, yshift=-16mm, align=center]:{\lossone\\$\mathcal{C}_{loss}^{(1)}$}}] {} (one);
        \draw[-stealth, ultra thick, Gray] (one) to node[pos=0.4, label={[xshift=-10mm, yshift=-5mm, align=center]:{\abo\\$\mathcal{C}_{abor.}^{\MZZ}$}}] {} (abo);

        \draw[-stealth, ultra thick, ForestGreen] (three) to node[midway, below] {\small \suc} (suc_p);

        \node[circle, fill=NavyBlue] (middle) at (4,0) {\color{white}$T_2$};
        \node[NavyBlue] at ($(middle)+(90:0.75)$) {$\mathcal{C}_{deco.}^{(t_{\rm cycle})}$};
        \node[scale=2.7, rotate=-90, white] (middle_cycle) at (4,0) {$\circlearrowright$};
    \end{tikzpicture}

    \caption{
    Schematic representation of the RUS subroutine in the presence of physical noise. Step {\bf (1)} corresponds to the emission of photons, step {\bf (2)} to the unitary transformation on the photonic modes $U(\varphi)$, and step {\bf (3)} to photonic detection. {\bf (a)} $\CZ$-SPOQC architecture.
    The cycle stops after a \losstwo\ detection event, in which case the phases of both spin qubits are lost, or after a \suc\ detection event, in which case the spin qubits get entangled through a $\CZ$ gate with lowered fidelity due to distinguishability. {\bf (b)} \fspoqc{} architecture.
    The cycle stops after a \suc\ detection event leading to a noisy $\MZZ$ measurement on the spin qubits.
    For both architectures, decoherence acts throughout the cycle, inducing a continuous dephasing depending on the decoherence time $T_2$. For both architecture, we fix a maximum amount of cycle repetitions $T_{\max}$. If the RUS subroutine did not stop before $T_{\max}$ cycles, the subroutine is aborted. In this case no entanglement is generated.}
    \label{fig:RUS}
\end{figure*}

We consider three types of physical noise processes: photon loss, photon distinguishability, and spin decoherence. These processes translate into computational noise models whose detailed description depends on the variant of the SPOQC architecture. However, the physical processes leading to the noise models do not change with the variant of the architecture, which allows for a fair performance comparison between codes and architecture variants. 

Furthermore, we limit ourselves to the effects of these three noise processes on the RUS subroutine alone and therefore on entangling operations in our architecture. As in \cite{Gliniasty2024Spin}, we assume complete and perfect SU(2) control on the emitters as well as perfect single-qubit measurements and initialization (See Section~\ref{sec:discussion_opening} for discussion about these assumptions).

Throughout this section, $\rho$ refers to the density matrix of the $n$-spin-qubit system. Single-qubit operators (resp.\ two-qubit operators) with a subscript $i$ (resp.\ $ij$) should be understood as acting on qubit $i$ (resp.\ $i$ and $j$). These subscripts are omitted when there is no ambiguity.

\subsubsection{Photon loss}
In photonic platforms, photon loss is the dominant source of noise. In the presence of photon loss, less than two photons can be detected, leading to \loss\ measurement patterns. This corresponds to heralded errors, whose error channels depend on the number of detected photons. If one photon is detected -- a \lossone\ event --, the phase of either one of the two spin qubits is erased, but we don't know which one. Up to some known single-qubit $\Z$ operations, this situation is described by the channel:
\begin{equation}
  \mathcal{C}_{loss}^{(1)}[\rho] = \frac{\mathcal{C}_a[\rho] + \mathcal{C}_b[\rho]}{2},
  \label{eq:c_rus1Photon}
\end{equation}
where $a, b$ denote the spin qubits and $\mathcal{C}_i[\rho]  =  \left( \rho + \Z_i \rho \Z_i \right)/2$ is the phase erasure channel \footnote{The error channel given by Eq. \eqref{eq:c_rus1Photon} is different from the one used in Ref.~\cite{Gliniasty2024Spin}, where 1-photon and no-photon detection patterns were associated to the same error channel, denoted $\mathcal{C}_{loss}^{(\infty)}[\rho]$ here. This new result corresponds to a finer understanding of the effect of loss in a RUS gate.}. If no photons are detected -- a \losstwo\ event --, the phases of both spin qubits are erased:
\begin{equation}
    \mathcal{C}_{loss}^{(\infty)}[\rho] = (\mathcal{C}_a \circ \mathcal{C}_b)[\rho].  \label{eq:c_rusNoPhoton}
\end{equation}
For both architecture variants, the RUS subroutine action after a \lossone\ event is the same as after a \bunch\ event: the cycle is repeated. Such events introduce additional noise, described by $\mathcal{C}_{loss}^{(1)}$, see Section~\ref{sec:discussion_opening} for details. \losstwo\ events, however, are treated differently in the two variants. For $\CZ$-SPOQC, a \losstwo\ events completely erases the phase of both qubits, such that the stabilizer measurement stored in the phase of the auxiliary qubit gets lost. Therefore there is no point in continuing the RUS subroutine, as the dephasing would entirely override a subsequent CZ gate. On  the other hand, in the $\MZZ$-SPOQC variant, the \losstwo\ event error channel commutes with the ideal measurement performed with \suc\ events. Therefore, the measurement outcome is unaffected by this error channel, and it is worth continuing the subroutine until a \suc\ event is recorded. We discuss this loss mechanism in more details in Appendix~\ref{app:loss}.

\subsubsection{Photon distinguishability}

Apart from the dual-rail-encoded qubit degree of freedom, photons used in the computation should be completely indistinguishable in order to interfere perfectly. In a realistic setting, that is not the case, and we model distinguishability through an additional internal degree of freedom $\ket{\xi}$ tensored with the qubit degree of freedom. The mean-wave packet overlap between photons emitted by spin qubits $a$ and $b$ is defined as the overlap between internal states $M_{ab}=|\langle\xi_a|\xi_b \rangle|^2$ and we drop the subscript for brevity. For simplicity, we assume that the overlap between emitters is fixed, see Section~\ref{sec:discussion_opening} for details. 
{In the case of a RUS $\MZZ$ operation, the physical error channel caused by photon distinguishability is:
\begin{equation}
    \mathcal{C}^{(\pm)}[\rho] = \left(1 - \frac{D}{2}\right) {\Pi}_{\pm}\rho{\Pi}_{\pm} + \frac{D}{2} {\Pi}_{\mp}\rho{\Pi}_{\mp},
\end{equation}
where $D=1-M$, the sign indicates the measurement outcome, and ${\Pi}_{\pm}=(\Id_a\Id_b\pm\Z_a\Z_b)/2$ is the projector onto the $\pm 1$ eigenspace of $\Z_a\Z_b$. These channels can be combined and rewritten, taking into account the measurement record: 
\begin{equation}
\begin{aligned}
    \mathcal{C}_{dist.}^{\MZZ}[\rho]&={\Pi}_{+}\rho{\Pi}_{+}\otimes\left[\left(1-\frac{D}{2}\right)\ketbra{0}{0}+\frac{D}{2}\ketbra{1}{1}\right]\\
    &+{\Pi}_{-}\rho{\Pi}_{-}\otimes\left[\left(1-\frac{D}{2}\right)\ketbra{1}{1}+\frac{D}{2}\ketbra{0}{0}\right],
\end{aligned}
\end{equation}
where we omitted the overall normalization for conciseness. The additional fictitious qubits represent the measurement record, with $\ket{0}$ (resp.\ $\ket{1}$) corresponding to measuring $+\Z\Z$ (resp.\ $-\Z\Z$). This channel thus corresponds to a measurement error occurring with probability $D/2$ (see Appendix \ref{app:distinguishability} for a detailed derivation of the result). 

In the case of a RUS $\CZ$ gate, the error channel leads to a unheralded $\Z\Z$ error on the two spin qubits:
\begin{equation}
\begin{aligned}
    \mathcal{C}_{dist.}^{\CZ}[\rho] &= \left(1 - \frac{D}{2}\right) \CZ_{ab} \rho \CZ_{ab} \\&\;+ \frac{D}{2} \Z_a\Z_b \CZ_{ab}\rho \CZ_{ab}\Z_a\Z_b.
\end{aligned}
\label{eq:distinguishability}
\end{equation}

Finally, distinguishability also introduces new detection patterns, where one photon is detected in each mode of a single dual-rail encoding, and no photon is detected in the other two modes, but these pattern detections have the same effect as the \bunch\ events described in Section~\ref{sec:RUS}. We thus include these patterns in the \bunch\ detection events.

\subsubsection{Spin coherence time} 
The spin degree of freedom of quantum emitters is highly sensitive to its surrounding environment, causing decoherence. This is for example the case of semiconductor quantum dots interacting with their surrounding nuclear spin bath, see e.g.~\cite{coste2023high}. Spin qubits decoherence is characterized by the spin coherence time $T_2$, which we assume to be much smaller than the spin relaxation time $T_1$. Under this assumption, the error channel corresponding to decoherence of emitter $i$ is:
\begin{equation}
    \mathcal{C}_{deco.,\,i}^{(t)}[\rho] = (1 - p_\Z(t)) \rho + p_\Z(t) \Z_i \rho \Z_i,
\end{equation}
with
\begin{equation}
    p_\Z(t) = \frac{1}{2}\left(1 - \exp[- \frac{t}{T_2}]\right).
\end{equation}
This result stems from a pure dephasing model and we refer the reader to Appendix C of Ref.~\cite{Gliniasty2024Spin} for a complete derivation. We assume that spin decoherence acts independently on all qubits during the implementation of RUS operations, and that the timescale of a RUS operation is much larger than that of single-qubit operations (initializations, gates, and measurements). This is motivated by the fact that the RUS gate requires a feed-forward operation, and it implies that the execution of a single RUS cycle takes a fixed amount of time $t_{\rm cycle}$, so that the applied channel is $\mathcal{C}_{deco.,\,i}^{(t_{\rm cycle})}$.

Since all of the error channels commute, it is possible to define an aggregate time $t_{RUS}=T\cdot t_{cycle}$, where $T$ is the number of cycles that the subroutine went through. This number is the total amount of time needed to terminate the RUS subroutine or more generally, the total amount of time we might want to allocate to the subroutine. Note that $t_{cycle}$ is more fine-grained and physical as it pertains directly to the cycle execution of the RUS gate, whereas $t_{RUS}$ is more abstract and can be interpreted as the time allocated to entangling layers in the computation.

\subsection{The RUS error channels}

To simulate a whole RUS subroutine, the error channels described in the previous section are composed at each cycle, and the number of cycles has to be capped.

\subsubsection{Single RUS gate} The noise channel for the entangling operations in the SPOQC architectures is determined by the repetition of the noisy RUS cycles described in Fig.~\ref{fig:RUS}. As RUS gates are independent, each RUS gate has a different noise channel that depends on its specific detection events history. For instance, denoting $(n_0,n_1,n_2,n_3)$ a measurement pattern with $n_i$ photons detected in mode $i$, a RUS $\CZ$ gate between emitters $a$ and $b$ with detection history $(0,2,0,0)\to(0,0,0,1)\to(1,0,1,0)$ will be represented by a channel
\begin{equation}
\mathcal{C}_{deco.}^{(t_{\rm cycle})}\circ\mathcal{C}_{dist.}\circ\mathcal{C}_{deco.}^{(t_{\rm cycle})}\circ\mathcal{C}_{loss.}^{(1)}\circ\mathcal{C}_{deco.}^{(t_{\rm cycle})}\circ\mathcal{I},
\end{equation}
with $\mathcal{C}_{deco.}^{(t)}=\mathcal{C}_{deco.,\,a}^{(t)}\circ\mathcal{C}_{deco.,\,b}^{(t)}$ (where we omitted subscripts $a$ and $b$ for clarity). For a RUS $\MZZ$ gate with detection history $(0,0,0,0)\to(0,1,1,0)$ we will have
\begin{equation} 
\mathcal{C}_{deco.}^{(t_{\rm cycle})}\circ\mathcal{C}^{(-)}\circ\mathcal{C}_{deco.,\,a,b}^{(t_{\rm cycle})}\circ\mathcal{C}_{loss.}^{(\infty)}.
\end{equation}
Assuming $T_2\ll T_1$, all these channels commute with one another, such that we can always rewrite a successful RUS $\CZ$ gate noise channel as
\begin{equation}
    \mathcal{C}_{deco.}^{(T\cdot t_{\rm cycle})}\circ\mathcal{C}_{loss}^{(N)}\circ\mathcal{C}_{dist.}^{\CZ},
    \label{eq:rus_channel_cz}
\end{equation}
and a RUS $\MZZ$ gate as 
\begin{equation}
    \mathcal{C}_{deco.}^{(T\cdot t_{\rm cycle})}\circ\mathcal{C}_{loss}^{(N)}\circ\mathcal{C}_{dist.}^{\MZZ}.
    \label{eq:rus_channel_mzz}
\end{equation}
where $(\cdot)^{(k)}$ represents the $k$-fold composition and the random variable $T$ associated to the number of cycles before a successful outcome was observed. The random variable $N$ can be expressed as $N=\sum_{i=1}^{T}k_i$ with $k_i\in\{0,1,\infty\}$ depending on whether a \bunch, a \lossone, or a \losstwo\ event was observed during cycle $i$, with the convention $\infty+k=\infty$, $k\geq0$. Finally, a failed RUS $\CZ$ gate will have a noise channel 
\begin{equation}
    \mathcal{C}_{deco.}^{(T\cdot t_{\rm cycle})}\circ\mathcal{C}_{loss}^{(\infty)},
    \label{eq:rus_channel_cz_fail}
\end{equation}
where $T$ is the number of cycles before a \losstwo\ event was observed (See Appendix~\ref{ap:simulation} for more details).

\subsubsection{Multiple RUS gates} 
To minimize the circuit depth of a QEC protocol, we can execute some entangling gates in parallel. As we need to wait for all of the RUS gates to terminate before going to the next step of the protocol, qubits involved in early-terminating RUS gates could idle for a while. Calling $T_i$ the number of cycle repetitions needed for RUS gate $i$ to terminate, the idling time for both qubits involved in this gate will be $(\max_{j}(T_j)-T_i)t_{\rm cycle}$. Consequently, we replace $T\cdot t_{\rm cycle}$ by $\max_{j}(T_j)\cdot t_{\rm cycle}$ in Eqs.~(\ref{eq:rus_channel_cz}), (\ref{eq:rus_channel_mzz}), (\ref{eq:rus_channel_cz_fail}).

Finally, as the number of qubits in the QEC code increases, the typical number of parallelly-executed RUS gates increases as well. Calling $L$ this number, we would have: 
\begin{equation}
    \mathbb{E}\left[\max_{1\leq j\leq L}(T_j)\right]
    \underset{L\to\infty}{\sim}
    -\log(L)/\log(1-p_{stop}),
\end{equation} with $p_{stop}$ the probability to exit the RUS subroutine at each cycle and and $\mathbb{E}[\cdot]$ the mathematical expectation. The noise induced by decoherence would therefore scale with the size of the encoding which would violate the assumptions of the threshold theorem. To circumvent this, we cap the number of cycles to $T_{\max}$ independently of the size of the encoding. If a RUS gate did not terminate before this cap, we record an \abo\ event. In this case, no entanglement could be generated. For simplicity, we treat this event as if entanglement were generated followed by a noise channel $\mathcal{C}_{abor.}^{\CZ}=\mathcal{C}_{loss}^{(\infty)}\circ\mathcal{C}_{dist.}^{\CZ}$ for $\CZ$-SPOQC and $\mathcal{C}_{abor.}^{\MZZ}=\mathcal{C}_{loss}^{(\infty)}\circ\left.\mathcal{C}_{dist.}^{\MZZ}\right|_{D=1}$ (a distinguishability channel with $D=1$, equivalent to losing the measurement record, followed by a \losstwo\ event channel). Essentially, we treat the \abo\ case as if a perfect RUS operation were applied, followed by a complete loss of information. This treatment is convenient for simulations as the detecting regions of the code do not have to be modified.

\section{Comparison between the surface code and the honeycomb Floquet code}

In~\cite{Gliniasty2024Spin}, we introduced the $\CZ$-SPOQC architecture and performed threshold simulations on the surface code for the three errors described above. We now expand this study to the planar honeycomb code~\cite{HH2022Boundaries}, which we can implement on both SPOQC variants, as we explain here, before analyzing the results of the simulations.

\subsection{Implementing the honeycomb Floquet code on the SPOQC architectures}

\begin{figure}
\centering
\begin{tikzpicture}
\node at (-1,0)
    {\includegraphics[width=0.45\textwidth]{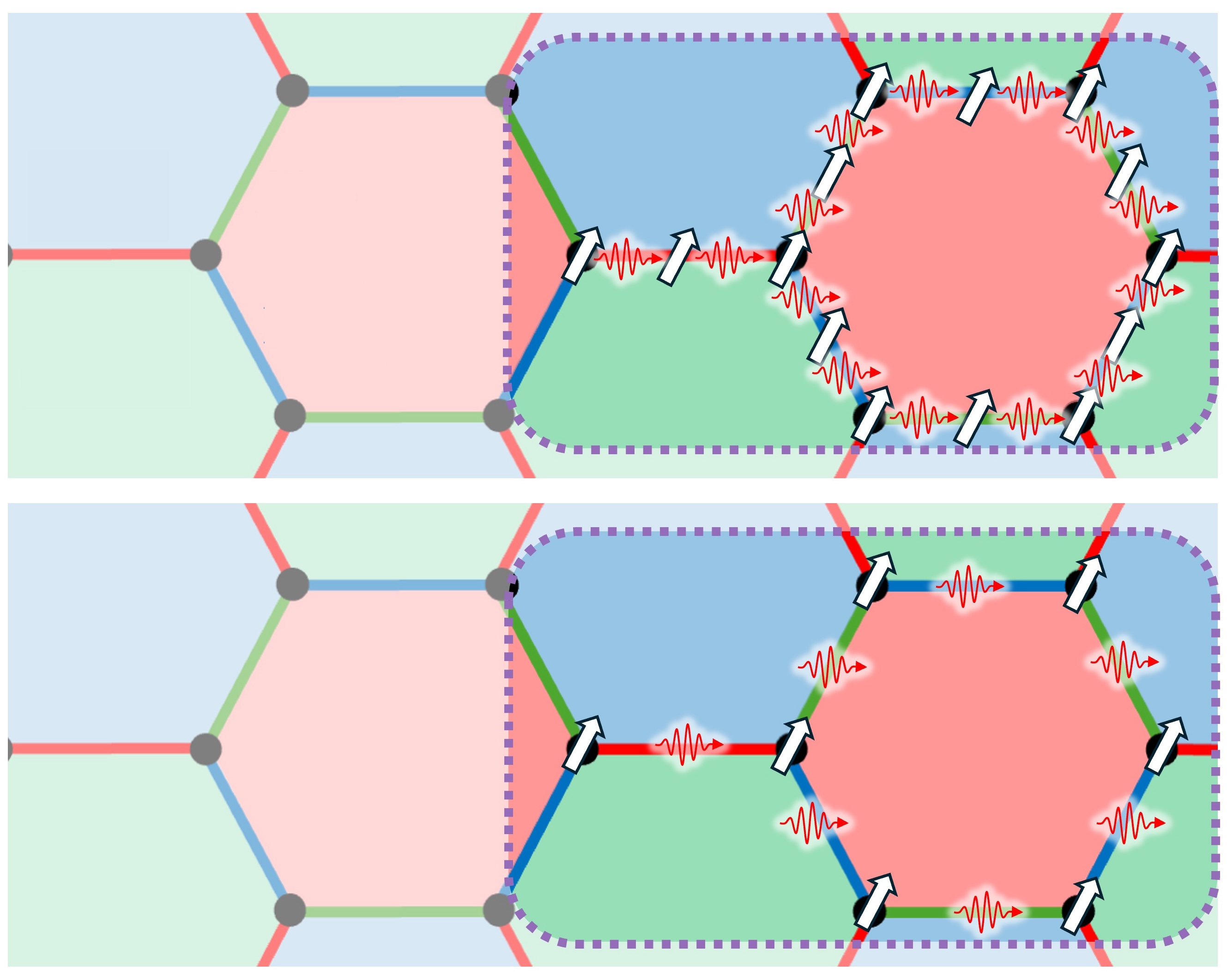}};
\begin{scope}[xshift=-0.62cm, yshift=-0.225cm]
    \node at (-2.2, 1.2) {\color{hcblue} \normalsize $\MZZ$};
    \node at (-2.2, 2.2) {\color{hcgreen}\normalsize $\MYY$};
    \node at (-3.5, 1.9) {\color{hcred} \normalsize $\MXX$};
    \node[align=center] at (-3.4, 0.7) {\bf Data \\\bf qubit};
    \draw[->, thick] (-3.2, 1.2) to (-3, 1.55);
\end{scope}
\node[fill=blue!70, opacity=0.7, rounded corners=7pt, minimum width=2.1cm, minimum height=0.7cm, label={center:{\color{white}\small $\CZ$-SPOQC}}] at (-1,2.5) {};
\node[fill=blue!70, opacity=0.7, rounded corners=7pt, minimum width=2.1cm, minimum height=0.7cm, label={center:{\color{white}\small \fspoqc}}] at (-1,-0.7) {};
\node at (-4.5,2.7) {$\textbf{(a)}$};
\node at (-4.5,-0.4) {$\textbf{(b)}$};
\end{tikzpicture}
\caption{ 
Honeycomb Floquet code defined by a sequence of $\MXX$, $\MYY$ and $\MZZ$ measurements on red, green and blue edges of a tri-colorable hexagonal lattice, with architecture layout for the $\textbf{(a)}$ $\CZ$-SPOQC and $\textbf{(b)}$ \fspoqc{} architectures. The white arrows represent quantum emitters and the wavy red arrows represent linear-optical RUS operations between emitters.}
\label{fig:honeycomb}
\end{figure}

The planar honeycomb code is a periodic dynamical code, aka Floquet code~\cite{HH2021Dynamically}, where data qubits are located on the vertices of a planar hexagonal tiling. The edges correspond to two-qubit non-destructive measurements on the incident data qubits. Such a tiling is edge-tri-colorable and a color is associated to each edge: red, green, and blue edges will correspond to $\MXX$, $\MYY$, and $\MZZ$ measurements, respectively. At a given time step, all the edges of a given color are measured in the corresponding measurement basis. We repeatedly perform the measurement sequence ``red-green-blue-red-blue-green"~\cite{GNM2022Benchmarking} which is compatible with the planar layout represented in Fig.~\ref{fig:honeycomb}.

To implement the honeycomb code on the $\CZ$-SPOQC architecture, data qubits are placed on each vertex of the hexagonal tiling, and an additional check qubit is placed on each edge, as depicted in Fig.~\ref{fig:honeycomb}(a). Each edge is associated to two RUS $\CZ$ gates, which allows to perform the syndrome extraction  shown on Fig.~\ref{fig:Syndrome}(a). On the \fspoqc{} architecture, the $\MZZ$ pair measurements are directly performed using the RUS gate described in the previous section, as depicted in Fig.~\ref{fig:Syndrome}(b). Using Hadamard gates $\Ha = \Ha_{\X\Z} = (\X + \Z)/\sqrt2$ and $\Ha_{\Y\Z} = (\Y + \Z)/\sqrt2$ gates on the spins, one can convert an $\MZZ$ measurement into an $\MXX$ or $\MYY$ measurement, thus obtaining all the measurements required to direclty perform the measurement sequence of the honeycomb code \footnote{More precisely, one must add $\Ha$ gates on each emitter before and after the RUS gate to perform an $\MXX$ measurement, and $\Ha_{\Y\Z}$ gates for $\MYY$ measurements.}.
This significantly reduces the numbers of spin qubits and RUS modules compared to the $\CZ$-SPOQC implementation of the honeycomb code. Indeed, in a honeycomb lattice, a vertex is shared by three faces and an edge is shared by two faces. There are thus $3/2$ times more edges than vertices for an infinite honeycomb lattice. Compared with the $\CZ$-SPOQC architecture, this translates in a $60\%$ reduction in the number of spin qubits, as we do not allocate spin qubits to the edges of the lattice, and it halves the number of RUS modules, since we only need one per edge for \fspoqc{}.

\begin{figure}
\centering
    \begin{tikzpicture}
    \node at (0,0)
        {\includegraphics[width=0.45\textwidth]{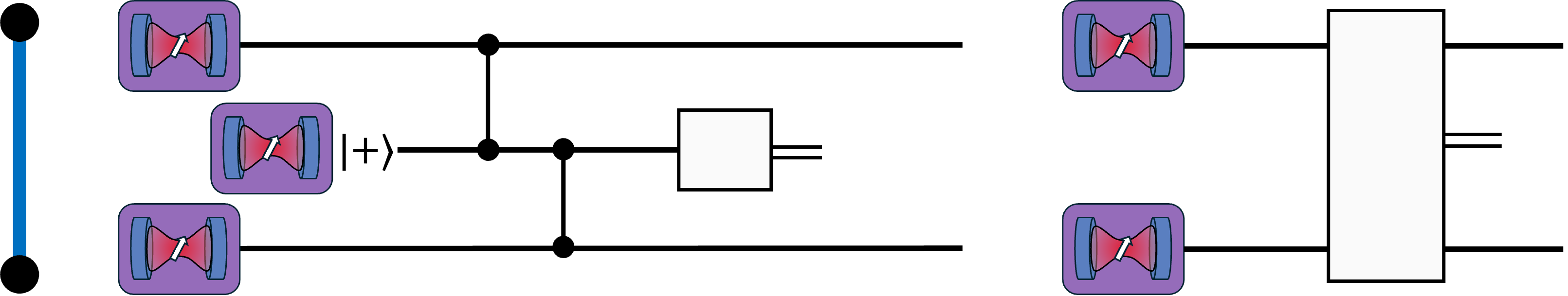}};
    \node at (-2.3, 1.05) {\small $\textbf{(a)} \ \CZ$-SPOQC};
    \node at (2.4, 1.05) { \small $\textbf{(b)}$ \fspoqc{}};
    \node[inner sep=1.5pt] at (-0.3, 0) {\scriptsize $\MX$};
    \node[inner sep=1.5pt] at (2.95, 0) {\scriptsize $\MZZ$};
    \node[inner sep=1.5pt] at (0.4, 0) {\scriptsize$m$};
    \node[inner sep=1.5pt] at (3.75, 0.05) {\scriptsize$m$};
\end{tikzpicture}
\caption{Implementation of an $\MZZ$ measurement for the $\textbf{(a)} \ \CZ$-SPOQC and $\textbf{(b)}$ \fspoqc{} (right) architectures, where $m$ denotes the measurement result.}
\label{fig:Syndrome}
\end{figure}

The interest of the \fspoqc{} architecture is however not restricted to dynamical codes. We can always design a syndrome extraction circuit based on two-qubit measurements and implement it with the \fspoqc{} architecture~\cite{moflic2024constant}. The pentagon tiling of the surface code~\cite{Gidney2023Pair} offers a good example of such a design, for which we also perform threshold simulations in this article. This enables us to compare two quantum error correcting codes, the surface code and the honeycomb code, with a quantum error correcting circuit tailored to each SPOQC variant: auxiliary-qubit-assisted syndrome extraction for the $\CZ$-SPOQC architecture, and pair-measurement-based syndrome extraction for $\MZZ$-SPOQC.

\begin{figure*}
    \centering
    \includegraphics{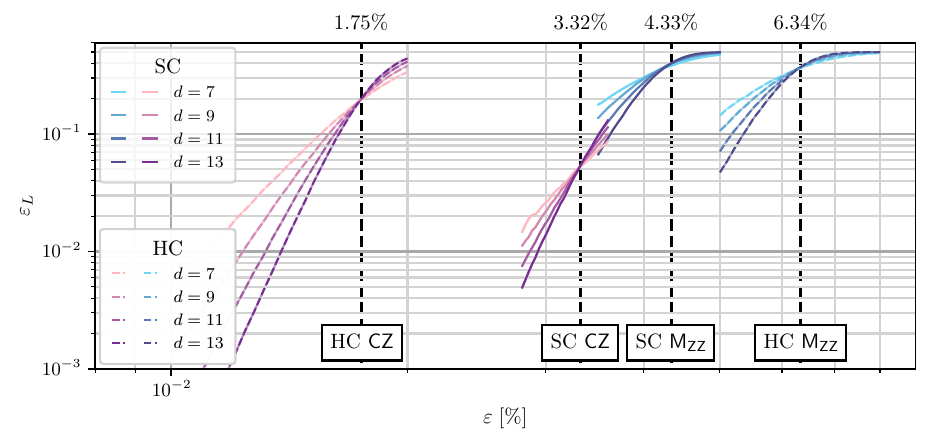}

    \caption{Logical error rate $\varepsilon_L$ as a function of photon loss $\varepsilon$ for different distances in the $\CZ$-SPOQC architecture (pink gradient) and $\MZZ$-SPOQC architecture (blue gradient). Thresholds values are reported on the top of the plots and marked by the black dashed lines. These curves are obtained by averaging out the mismatch between the measure of the logical observable $\X$ obtained through a data readout at the end of the QEC cycles, and the predicted observable by the MWPM decoder \cite{higgott2023sparse}.}
    \label{fig:honeycomb_p_erasure}
\end{figure*}

\subsection{Memory threshold comparisons for individual noise mechanisms}

\begin{table}[h]
\small
\begin{center}
\begin{tabular}{|c |c ||c| c | c|} 
 \hline
 Code & SPOQC & $\varepsilon$ & $D$ & $t_{\rm RUS} / T_2$ \\
 \hline
Surface & $\CZ$ & 3.3\% & 2.4\% & 2.5\% \\
\cline{2-5} code & $\MZZ$ & 4.3\% & * & 1.2\% \\
 \hline
Honeycomb & $\CZ$ & 1.8\% & 1.9\% & 0.6\% \\
 \cline{2-5} Floquet code & $\MZZ$  & 6.3\% & 2.4\% & 2.3\% \\ 
 \hline
\end{tabular}
\end{center}
\caption{Summary of the achievable individual thresholds rounded to the second significant figure for the surface and  honeycomb Floquet codes and for the SPOQC (native $\CZ$ operation) and \fspoqc~(native $\MZZ$ operation) architectures. Surface code with $\CZ$ results comes from Ref.~\cite{Gliniasty2024Spin}, surface code with $\MZZ$ exploits the syndrome extraction circuit from Ref.~\cite{Gidney2023Pair}, and honeycomb code with both $\CZ$ and $\MZZ$ native operations were implemented using the strategy described above. Surface code with $\MZZ$-SPOQC is insensitive to distinguishability error only.}
\label{table:thresholds}
\end{table}

To compare both architectures, we rely on the computation of \emph{fault-tolerant thresholds}. For a given abstract physical error parameterized by $p$ (typically the physical error probability or a quantity growing with that probability), the threshold is the value $p_{th}$ under which increasing the size of the code yields increasingly better protection of the logical information. We can thus compare architectures through the various thresholds they offer: the higher the threshold, the more effective the architecture is at correcting this type of error. In this article, we focus on the threshold specific to an $\X$-type quantum memory, i.e., the preparation and protection of a logical $\ket{+_L}$ state (the $\Z$-type quantum memory actually exhibits consistently similar or larger thresholds). Protecting a quantum memory is the simplest operation required by a fault-tolerant quantum computer, and is therefore an important benchmark for any architecture.

Table~\ref{table:thresholds} summarizes the different threshold values obtained for $\CZ$-SPOQC and \fspoqc{} given different photonic error models corresponding to photon loss, distinguishability and spin decoherence. Distinguishability and decoherence error models are respectively parameterized by $D=1-M$ where $M$ is the mean wave-packet overlap between photons, and the ratio $t_{\rm RUS}/T_2$ between the time $t_{\rm RUS}$ for the RUS subroutine to terminate and the spin coherence time $T_2$. To evaluate the thresholds, we numerically estimate the logical error rates, i.e.~the probability of logical information corruption, of honeycomb codes of various lattice sizes (parameterized by the distance $d$ of the code) and plot it as a function of the physical error parameter. The crossing point between the curves corresponding to the two largest simulated distances gives an estimation of the fault-tolerant threshold. See Appendix \ref{ap:simulation} for more details. 

\subsubsection{Loss threshold}
Fig.~\ref{fig:honeycomb_p_erasure} shows the evolution of the logical error rate as a function of the loss $\varepsilon$ and $d$, for given codes and architectures. A high loss threshold is crucial in photonic architectures since photon loss is the most important source of noise in photonic platforms: state-of-the-art photonic experiments reach a photon loss of about $\varepsilon \approx 30\%$ \cite{highefficiencyJWP}. Here, with the honeycomb code implemented on the \fspoqc{} architecture, we obtain a photon loss threshold of $\varepsilon_{th} \approx 6.3\%$. This is significantly larger than what is achieved with a surface code on the $\CZ$-SPOQC architecture, for which the threshold is $\varepsilon_{th} \approx 3.3\%$ (Note that this value is higher than the $2.8\%$ value reported in \cite{Gliniasty2024Spin}; this is due to a finer understanding of the loss mechanism in this work compared to \cite{Gliniasty2024Spin}, where any loss event was assumed to dephase both emitters). 

The threshold increase between the two architectures can be explained in the following way. For RUS $\CZ$ gates, any \loss\ event carries a probability of completely losing the phase of the auxiliary qubit involved in the RUS gate, and thus of losing the stabilizer measurement. On the other hand, the \loss\ channels commute with the $\MZZ$ and hence does not change the measurement result. With RUS $\MZZ$ measurements, we thus repeat the subroutine even if we have a complete dephasing of both emitters. Importantly, this means that we can always recover the measurement outcome by using sufficiently many trials. In other words, the success probability of obtaining an unambiguous measurement outcome is 1 (provided no other errors are considered). We believe that this is the main source of threshold improvement of the \fspoqc{} architecture when tailored to Floquet codes. 

For completeness, we also implement the surface code on the \fspoqc{} architecture. To do so, we need syndrome extraction circuits that only involve pair measurements~\cite{CBDH2020Optimization, Gidney2023Pair}. Here, we use the planar pentagon tiling layout~\cite{Gidney2023Pair}. Our simulations show that, for the surface code as well, a larger loss threshold is obtained with the $\MZZ$-SPOQC architecture, but the improvement is not as important as for the honeycomb code.

\begin{figure}[!t]
\centering
\begin{subfigure}{0.48\textwidth}
    \includegraphics{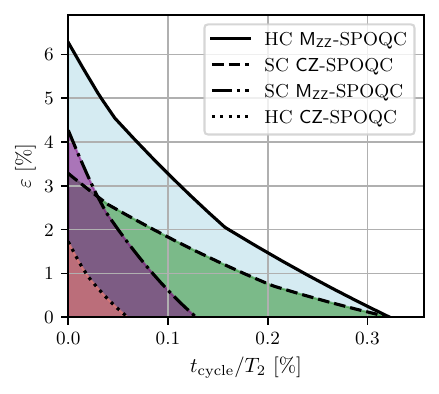}
    \caption{Fault-tolerant regions}
\end{subfigure}
\begin{subfigure}[][4cm][c]{0.45\textwidth}
    \small
    \begin{tikzpicture}
        \node [shape=rectangle, align=left](table1) at (-0.2,-6) {\small
            \begin{tabular}{ |c|c||c| } \hline
                Code & SPOQC  & Area\\
                \hline
                HC & $\MZZ$ & $7.7\times 10^{-5}$\\
                \hline
                SC & $\CZ$ & $4.2\times 10^{-5}$\\
                \hline
                SC & $\MZZ$ & $2.2\times 10^{-5}$\\
                \hline
                HC & $\CZ$ & $0.4\times 10^{-5}$\\
                \hline
            \end{tabular}
        };

        \begin{scope}[xshift=0.75cm]
            \draw[-stealth] (1.75,-7.2) to[bend right] node[midway, right] {$\times 5.2$} (1.75,-6.65);
            \draw[-stealth] ($(1.75,-7.2)+(0,0.6)$) to[bend right] node[midway, right] {$\times 1.9$} ($(1.75,-6.65)+(0,0.6)$);
            \draw[-stealth] ($(1.75,-7.2)+(0,1.2)$) to[bend right] node[midway, right] {$\times 1.8$} ($(1.75,-6.65)+(0,1.2)$);
            \draw[-stealth] (3,-7.2) to[bend right] node[midway, right] {$\times 17.9$} ($(3,-6.65)+(0,1.2)$);
        \end{scope}
\end{tikzpicture}
\caption{Area comparison}
\end{subfigure}

\caption{(a) Fault-tolerant areas shaded in blue for pairs of codes and architectures. These areas bound the region of space where increasing the size of the encoding yields better protection of the logical information in the presence of photon loss and quantum emitter decoherence. (b) Comparison of fault-tolerant areas. HC (resp.\ SC) abbreviates honeycomb code (resp.\ surface code). The different combination of codes and architecture overlap neatly such that comparing the areas of the fault-tolerant regions gives a proxy of relative performances. The honeycomb code in $\MZZ$-SPOQC performs $1.8$ times better than the surface code in $\CZ$-SPOQC, $3.5$ times better than the surface code in $\MZZ$-SPOQC and $17.9$ times better than its counterpart in the $\CZ$ based implementation.}
\label{fig:2d_ft_regions}
\end{figure}

\subsubsection{Decoherence threshold}
\label{sec:deco_thr}

For decoherence, we found similar thresholds for the honeycomb code with \fspoqc{} and the surface code with $\CZ-$SPOQC, and slightly inferior results for the pair-measurement syndrome extraction circuits of the surface code with \fspoqc{}. The poorer performances of the surface code implemented on \fspoqc{} can mostly be explained by two factors. First, the higher depth of the syndrome extraction circuit for the pentagon-tiled surface code allows more time for spin decoherence. The pentagon-tiled surface code also uses 2 auxiliary qubits instead of one, so more qubits are subject to decoherence. Logical error rates curves can be found in Appendix \ref{ap:other_thr_plots}.

\subsubsection{Distinguishability threshold}
\label{sec:dist_thr}

For distinguishability errors, comparison is more tedious. Indeed in the presence of two-qubit measurement errors, i.e. errors induced by distinguishability, it is known that the honeycomb Floquet code can exhibit hyper-edge error mechanisms \cite{setiawan2024tailoringdynamicalcodesbiased, fahimniya2024faulttoleranthyperbolicfloquetquantum}. In this setting, decoding with minimum-weight perfect matching (MWPM) can be possible if one is able to decompose the hyper-edges into edges, paying a price in decoding accuracy. However, it does not seem possible to do so in our context since we simulate only this specific error mechanism, and MWPM fails to find adequate matchings. We thus need to decode using a more general but slower decoder such as belief-propagation with ordered statistics post-processing (BP+OSD) \cite{roffe_decoding_2020}, which requires extensive computation times. With this decoder, the threshold found for distinguishability is around 2.4\%, which is comparable to that of the surface code in the $\CZ$-SPOQC architecture. This result might seem surprising, since distinguishability errors have more impact in the $\CZ$-based architecture, as they simultaneously flip the measurement outcome and introduce noise on the data qubits. We believe that the threshold for the honeycomb code on the $\MZZ$-based architecture is actually higher, and that we observe this value because we stop the OSD post-processing at order 5, to ensure a tractable decoding time. To support this intuition, we performed simulations for the honeycomb code where we added, on top of distinguishability errors, a small one-qubit depolarizing noise after each layers of two-qubit measurements. This additional fictitious noise is enough to decompose hyper-edge mechanisms and use MWPM. In this setting, the threshold was found to be around 2.8\% which, in anticipation to Figure~\ref{fig:3d_ft_regions} in Section~\ref{sec:3d_ft}, seems to be the accurate intersection point for the green and red curves, i.e. the threshold lines in the planes loss-distinguishability and decoherence-distinguishability, respectively. However, we report here the value of 2.4\% that we obtained in a rigorous way using BP+OSD.

Logical error rates curves can be found in Appendix \ref{ap:other_thr_plots}.

A surprising result is that of the pentagon tiling surface code when considering a distinguishability-only error model. The numerical simulations show that the logical error rate is always zero for any value of distinguishability. In other words, the logical error rate is insensitive to distinguishability when it is the only type of errors in the system. The reason for that is that we can retrieve both the logical $\Z$ and $\X$ operators of the surface code only with single-qubit measurement results: two-qubit measurements do not flip the logical observables in that case. This can be made clear by observing the detector error model output by the circuit under this noise model. The second thing to note is that distinguishability errors affect two-qubit measurements, but single-qubit measurements do not rely on two-photon interference and are thus insensitive to distinguishability errors. Therefore, we can perform reliably the single-qubit measurements required to keep track of the logical observables and this explains why logical information is insensitive to this specific noise model. However, this does not mean that the pentagon tiling surface code is insensitive to distinguishability errors in combinations of other type of errors. 

\subsection{Fault-tolerant regions}

\begin{figure*}[t]
    \centering
    \includegraphics[scale=0.95]{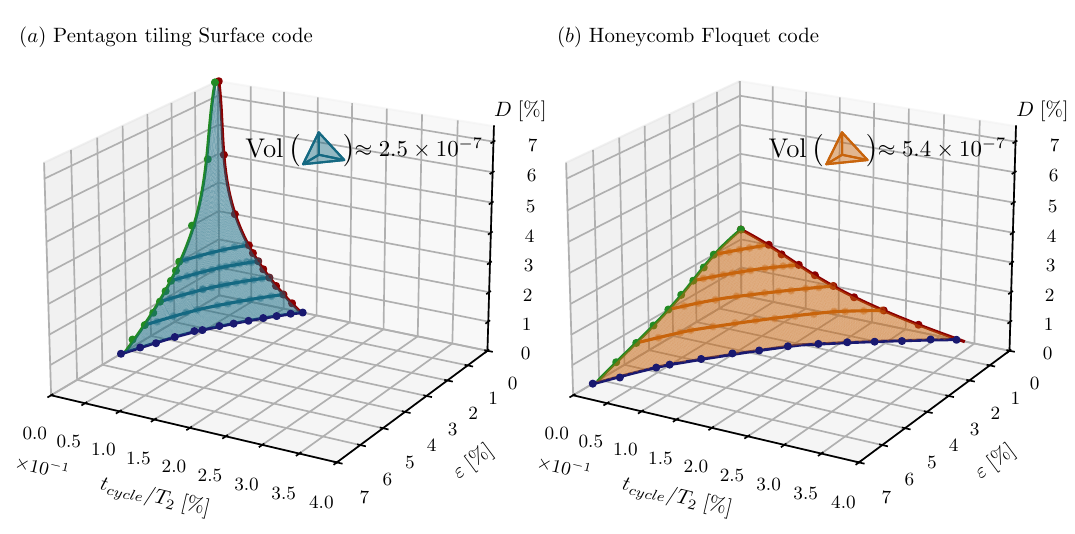}
    \caption{Fault-tolerant surfaces for $(a)$ the pentagon tiling of the surface code and $(b)$ the Honeycomb Floquet code in loss, decoherence, distinguishability space. Below these surfaces, increasing the number of qubits in the encoding yields increasingly better protection of the logical information in the presence of loss, distinguishability and decoherence simultaneously. Solid lines are polynomial fit interpolating the data points. Surfaces are obtained by linear grid interpolation of the solid lines. Volumes $\mathrm{Vol}(\cdot)$ are computed respectively to the unit cube of the space (see Appendix \ref{ap:derivation_ft_regions} for explanations on the computation). Overall, comparing the volumes of the fault-tolerant regions, the Honeycomb Floquet code performs $\sim 2.2$ times better than the pentagon-tiling. Although the pentagon tilling is insensitive to distinguishability errors alone, high distinguishability yields poorer performances when the code is subjected to additional loss and decoherence errors.}
    \label{fig:3d_ft_regions}
\end{figure*}

In real world scenarii, computational errors come from multiple error sources at once and we must therefore study the performances of the architectures in this context. It is all the more relevant to study joint errors given the unexpected behavior of the pentagon-tiled surface code with respect to distinguishability errors.

The study of joint errors relies on the computation of threshold hypersurfaces in the error space. They are the natural generalization of thresholds in the sense that they bound the region of error space where increasing the size of the encoding yields better protection of logical information (see Appendix \ref{ap:derivation_ft_regions}} for details on the derivation of these surfaces).

\subsubsection{Fault-tolerant areas in the loss-decoherence plane}

Fig.~\ref{fig:2d_ft_regions} shows the fault-tolerant areas obtained for perfectly indistinguishable photons. The honeycomb code implemented with pair-measurements (Fig.~\ref{fig:2d_ft_regions}(d)) again outperforms the surface code implemented with $\CZ$ gates (Fig.~\ref{fig:2d_ft_regions}(c)). As the FT area of the honeycomb code $\MZZ$-SPOQC contains that of the surface code $\CZ$-SPOQC (except for a very small region in the low-loss around-decoherence-threshold regime), we can directly compare the areas to estimate the relative performance of the two codes. Doing so, we obtain a $84\%$ increase in performance for the honeycomb code code in its native pair-measurement implementation.

In the loss-decoherence plane, codes implemented with their native gate set offer the best performances overall in terms of fault-tolerant area, although the surface code shows a small tradeoff between implementations. It is better to use the $\CZ$-based implementation of the surface code for $t_{cycle}/T_2\gtrsim 0.03\%$, but the pair-measurement implementation performs better below this bound. The better performance of the native implementation supports the idea of co-design developed in \cite{chan2024tailoring, cain2024correlated, setiawan2024tailoringdynamicalcodesbiased}, i.e. of jointly optimizing parts of the FTQC stack for increase performances.

\subsubsection{\texorpdfstring{Fault-tolerant volumes for the surface code and honeycomb code in $\MZZ$-SPOQC}{}}
\label{sec:3d_ft}

The comparison of the two-dimensional fault-tolerant regions in the loss-decoherence plane for the honeycomb code in $\MZZ$ and $\CZ$-SPOQC should be enough to draw conclusions in terms of relative code performances in the loss-decoherence-distinguishability space. Indeed, with 
lower distinguishability thresholds than
the honeycomb code with pair-measurements implementation and with thresholds curves exhibiting consistent convexity, the surface code $\CZ$-SPOQC and the honeycomb code $\MZZ$-SPOQC should exhibit smaller fault-tolerant volumes in the loss-decoherence-distinguishability space.

However, the comparison is more tedious with the pair-measurement implementation of the surface code due to its insensitivity to distinguishability errors alone. Fig.~\ref{fig:3d_ft_regions} shows the fault-tolerant volumes in the loss-decoherence-distinguishability space for the surface code (Fig.~\ref{fig:3d_ft_regions}(a)) and honeycomb code (Fig.~\ref{fig:3d_ft_regions}(b)) implemented with pair-measurements. Both volumes are obtained by computing thresholds surfaces using the MWPM decoder \footnote{Since the honeycomb code with pair-measurements exhibit non-decomposable hyperedges, we count a logical error whenever the syndrome cannot be decoded using MWPM.}. We see that the pentagon tiling of the surface code cannot sustain its error correcting capabilities in the presence of a high distinguishability. Indeed, the correctable volume for distinguishability values past $7.5\%$ (which is the highest distinguishability we simulated for Fig.~\ref{fig:3d_ft_regions}(a)) represents less than 1.2\% of the overall correctable volume. Although the pentagon tiling is still effective for distinguishability past $2.4\%$, the maximum tolerable loss and decoherence are respectively around $1.2\%$ and $0.03\%$. Overall the honeycomb code performs 2.2 times better than the pentagon tiling of the surface code in terms of correctable volume.

\section{Discussion}
We have carried out simulations with a noise model corresponding to the three dominant types of errors for a hybrid spin-photon platform exploiting the SPOQC architecture. These choices allow us to gain valuable insights on the performances of our architectures. It does however require some simplifications in the modelling of the real hardware. In this section, we provide an analysis of these choices, motivate them in greater details and suggest some improvements related to the error model.

A first avenue of extension for our work is the refinement of the physical error model. For instance, we do not consider the indirect effect of spin decoherence on photon distinguishability. Indeed, spin decoherence is usually induced by spectral jitters of the energy levels that also cause photon to be distinguishable. There's thus an interplay between photon distinguishability and spin decoherence that could be taken into account in more accurate simulations~\cite{TimeDepHOM_Thoma_2016, he_coherently_2019}.
However, the recent development of high-performance quantum-dot sources with almost-time-independent highly indistinguishable photons \cite{maring2024versatile} validates the simplification we made. In addition, the error channel for decoherence is extracted from a pure dephasing model. Although this model is analytically tractable, it is quite pessimistic as to the actual effect of decoherence in the system. Departing from the pure relaxation model and modeling more precisely the interaction between the quantum dot and the surrounding Overhauser field (see for instance \cite{coste2023high}) created by the nuclear bath could yield a better resilience of the SPOQC architectures to decoherence.

Extending the error model to include other relevant spin-optical error mechanisms could provide deeper insights into the tailoring of the SPOQC architectures.
Multi-photon emission, experimentally characterized by the intensity correlation $g^{(2)}(0)$ \cite{Wein:2021ahe}, should be taken into account in the future. In the absence of photon loss, multi-photon emission leads to heralded errors, since strictly more than two photons are detected during a RUS gate. We expect its fault-tolerance to be at least as large as for photon loss. 
In the presence of loss errors, multi-photon emission would require further study and is harder to anticipate.

Another important extension of the error model would be to include single-qubit errors. In this study, we assumed perfect SU(2) control on the emitters' spin qubits. This can be achieved by means of the spin precession around the axis of a low magnetic field, along with optical pulses applied to the emitter \cite{bodey_optical_2019, coste2023high}. These noisy operations should be taken into account in the design of the RUS subroutine. Indeed, \bunch\ events do not modify the state of the emitters as long as we apply a $\Z$ correction. Therefore, in the RUS cycle of Fig.~\ref{fig:RUS}, the noise channel associated to \bunch\ events should actually be the channel associated to a $\Z$ gate. Similarly, for \suc\ events in the $\CZ$-SPOQC architecture, the noise channel should be further composed with that of the $\S_a\S_b$ operation. The RUS subroutine should therefore be made such that we can track the corrections. We should apply them at the end of the subroutine to avoid unnecessary noise accumulation. Similarly, noiseless spin initialization and measurement are assumed. They both can be achieved through emission and measurement of the emitted photon. Single-qubit measurement errors should be considered even for pair-measurement extraction circuits since boundary stabilizers rely on some single-qubit measurements.

A second important avenue would be to improve the way we deal with \abo\ and \lossone\ event. For this study, we treat \abo\ events as maximum information loss after a perfect RUS operation. This is a strong pessimistic approximation, but it is very convenient for simulations. Indeed, since we still perform a perfect operation in the \abo\ case, the structure of the code and the detecting regions are unchanged. The proper way to deal with \abo\ events would be to dynamically adapt the extraction circuit to the changes of the detecting regions due to some entanglement being absent. Similarly, we could set up an upper bound on the number of \lossone\ events we can accept throughout a RUS subroutine. Above such bound, we stop the cycle since too much noise would have been introduced anyway. Such scenario would be treated in the same way as \abo\ events.  More generally, these improvements fall into the category of adaptivity, see \cite{Gliniasty2024Spin} Section 6 for an account of adaptivity within the original SPOQC architecture.

\section{Conclusion}
\label{sec:discussion_opening}

By tailoring the native operations of the spin-optical quantum computing architecture to the honeycomb Floquet code, we observe an improved photon loss threshold of 6.3\%, compared to the 3.3\% obtained withing the $\CZ$-SPOQC architecture, along with significant reduction in resource overheads.
This demonstrates concretely in the case of the SPOQC architectures that it is beneficial to co-design and seek compatibility between quantum error-correcting codes, syndrome extraction methods, and a hardware's native gates. We believe that this observation should also apply to other technological platforms, and we expect that extending the approach to include further building blocks of a fault-tolerant quantum computer, in particular decoding methods and logical gate implementations, can also prove fruitful, as suggested in ~\cite{bluvstein2024logical, cain2024correlated} in the context of cold atom platforms.

The idea of ``floquetifying'' quantum error-correcting codes has already been investigated in specific instances (e.g.~\cite{BLN2024Unifying, TMK2023Floquetifying, alam2024dynamical}). Our results lend concrete evidence to these efforts that the use of Floquet codes can lead to significant performance improvements.

Interestingly, we also observe an improvement in the photon loss threshold when implementing the surface code with a syndrome extraction circuit exploiting native two-qubit measurements,
which suggests that searching for efficient syndrome extraction circuits based on native two-qubit measurements is a noteworthy research avenue. While those available for the surface code have recently been improved~\cite{GransSamuelsson2024improvedpairwise}, it would be useful to transpose these results to other codes, such as high-encoding-rate quantum low-density parity-check codes.

Our work not only establishes the superiority of the honeycomb Floquet code over its static counterpart in the appropriate spin-optical quantum computing architecture but also highlights the potential of Floquet codes in advancing fault-tolerant quantum computing on photon-mediated platforms.

\section*{Acknowledgments}
We thank Nicolas Maring, Rawad Mezher, and Noah Shofer for their careful revision of this manuscript.
We thank Craig Gidney for the development of the Stim package and for open-sourcing the quantum circuits that we used to carry out our simulations.
This work has been co-funded by the European Commission as part of the EIC accelerator program under the grant agreement 190188855 for the SEPOQC project, by the Horizon-CL4 program under the grant agreement 101135288 for the EPIQUE project, by the PROQCIMA and TUF-TOPIQC programs within the French National Quantum Strategy (France 2030), and by the CIFRE grant n°2022/0532.
G. V-R acknowledges support from the Naquidis Innovation Center for Disruptive Quantum Products. 

\appendix

\section{Effect of the linear-optical interferometers in the noiseless and noisy cases}

In this appendix, we derive the different expressions for the spin operations performed through the detection of photons in the different patterns described in the main text. We also derive the noise channels associated with photon distinguishability and photon loss. The spin-optical architectures considered in our work rely on entangling spin qubits with photonic qubits. Gates are then indirectly applied on the spin qubits by means of operations, including measurements, on the spin-entangled photonic qubits, as depicted in Fig.~\ref{fig:RUS} of the main text.

\subsection{Noiseless case}
\label{app:interferometer}

In the following, we describe the different possible detection patterns and their subsequent effects on spins $a$ and $b$, assuming they are initially in the generic two-qubit states $\ket{\phi_{ab}}=\alpha\ket{0_{\rm s}0_{\rm s}}+\beta\ket{0_{\rm s}1_{\rm s}}+\gamma\ket{1_{\rm s}0_{\rm s}}+\delta\ket{1_{\rm s}1_{\rm s}}$.
The photonic qubits are dual-rail encoded: two optical modes per qubit are used and the logical `zero' state corresponds to having a photon in the first mode and no photon in the second mode, while the logical `one' state corresponds to a photon in the second mode and no photon in the first mode:
\begin{equation*}
    \ket{0_{\rm ph}} = \ket{\overline{10}}, \quad \ket{1_{\rm ph}}=\ket{\overline{01}},
\end{equation*}
where we use a horizontal bar to indicate Fock states in optical modes.
In the following, $|\overline\emptyset\rangle$ denotes the vacuum state (in all relevant optical modes) and $\hat{a}^\dagger_k$ denote the creation operator for mode $k \in \{0,1,2,3\}$. The two modes of the dual-rail qubit emitted by the first (respectively second) spin are taken to be modes 0 and 1 (respectively modes 2 and 3) of the interferometer. The operator $\hat{a}^\dagger_k$ acting on the vacuum state $|\overline\emptyset\rangle$ creates a photon in mode $k$ (we will omit the vacuum state in the following for clarity). The state of the system after the emission process can thus be written as: 
\begin{equation*}
\begin{aligned}
    &\alpha \ket{0_{\rm s}0_{\rm s}}\hat{a}_0^\dagger \hat{a}_2^\dagger + \beta \ket{0_{\rm s}1_{\rm s}} \hat{a}_0^\dagger\hat{a}_3^\dagger \\&+ \gamma \ket{1_{\rm s}0_{\rm s}}\hat{a}_1^\dagger \hat{a}_2^\dagger +\delta \ket{1_{\rm s}1_{\rm s}} \hat{a}_1^\dagger\hat{a}_3^\dagger
\end{aligned}
\end{equation*}
The optical modes then enter the interferometer from Fig.~\ref{fig:interferometer} of the main text, described by the matrix
    \begin{align}
U(\varphi)&=
\frac{1}{2}\begin{pmatrix}
    1 & 1 & 1 & -1 \\
    1 & 1 & -1 & 1 \\
    e^{i\varphi} & -e^{i\varphi} & 1 & 1 \\
    -e^{i\varphi} & e^{i\varphi} & 1 & 1
\end{pmatrix},
\end{align}
indicating how the creation operators of the four input modes are transformed.
The choice of the phase $\varphi$ allows to choose between two variants of the architecture.
After going through the interferometer, the system is in the state
\begin{equation*}
\begin{aligned}
    \ket{s_{\text{out}}}&=\alpha \ket{0_{\rm s}0_{\rm s}}\hat{a}_{0,\text{out}}^\dagger \hat{a}_{2,\text{out}}^\dagger + \beta \ket{0_{\rm s}1_{\rm s}} \hat{a}_{0,\text{out}}^\dagger\hat{a}_{3,\text{out}}^\dagger \\&+ \gamma \ket{1_{\rm s}0_{\rm s}}\hat{a}_{1,\text{out}}^\dagger \hat{a}_{2,\text{out}}^\dagger +\delta \ket{1_{\rm s}1_{\rm s}} \hat{a}_{1,\text{out}}^\dagger\hat{a}_{3,\text{out}}^\dagger
\end{aligned}
\end{equation*}
where
\begin{align*}
    \hat{a}_{0,\text{out}}^\dagger &= (\hat{a}_0^\dagger + \hat{a}_1^\dagger + e^{i \varphi} \hat{a}_2^\dagger - e^{i \varphi} \hat{a}_3^\dagger)/2,\\
    \hat{a}_{1,\text{out}}^\dagger &= (\hat{a}_0^\dagger + \hat{a}_1^\dagger - e^{i \varphi} \hat{a}_2^\dagger + e^{i \varphi} \hat{a}_3^\dagger)/2,\\
    \hat{a}_{2,\text{out}}^\dagger &= (\hat{a}_0^\dagger - \hat{a}_1^\dagger + \hat{a}_2^\dagger + \hat{a}_3^\dagger)/2,\\
    \hat{a}_{3,\text{out}}^\dagger &= (-\hat{a}_0^\dagger + \hat{a}_1^\dagger + \hat{a}_2^\dagger + \hat{a}_3^\dagger)/2.
\end{align*}

Depending on the detection patterns of the photons, the spin qubits are projected onto different states. To synthesize the outcomes, for each detection pattern, we list the amplitudes of every spin-qubit state in the canonical basis in Table~\ref{tab:statesout}: the outcome spin-qubit state for every detection pattern can be read on the table's rows.
\begin{table}
    \centering
    \begin{tabular}{|c|c|c|c|c|}
    \hline  
         & $\alpha \ket{0_{\rm s}0_{\rm s}} $&$\beta \ket{0_{\rm s}1_{\rm s}} $ &$\gamma \ket{1_{\rm s}0_{\rm s}}$ &$\delta \ket{1_{\rm s}1_{\rm s}}$  \\
         \hline 
        $ \hat{a}_0^\dagger \hat{a}_0^\dagger $&  $1$ & $-1$ & $1$ & $-1$ \\
        $ \hat{a}_1^\dagger \hat{a}_1^\dagger $&  $-1$ & $1$ & $-1$ & $1$ \\
        $ \hat{a}_2^\dagger \hat{a}_2^\dagger $&  $e^{i \varphi}$ & $e^{i \varphi}$ & $-e^{i \varphi}$ & $-e^{i \varphi}$ \\
        $ \hat{a}_3^\dagger \hat{a}_3^\dagger $&  $-e^{i \varphi}$ & $-e^{i \varphi}$ & $e^{i \varphi}$ & $e^{i \varphi}$ \\
        $\hat{a}_0^\dagger \hat{a}_1^\dagger$ & $0$ & $0$ & $0$ &$0$ \\
        $\hat{a}_2^\dagger \hat{a}_3^\dagger$ & $0$ & $0$ & $0$ &$0$ \\
        \hline
        $\hat{a}_0^\dagger \hat{a}_2^\dagger$ & $1+e^{i \varphi}$ & $1-e^{i \varphi}$ & $1-e^{i \varphi}$ &$1+e^{i \varphi}$ \\
        $\hat{a}_0^\dagger \hat{a}_3^\dagger$ & $1-e^{i \varphi}$ & $1+e^{i \varphi}$ & $1+e^{i \varphi}$ &$1-e^{i \varphi}$ \\
        $\hat{a}_1^\dagger \hat{a}_2^\dagger$ & $1-e^{i \varphi}$ & $1+e^{i \varphi}$ & $1+e^{i \varphi}$ &$1-e^{i \varphi}$ \\
        $\hat{a}_1^\dagger \hat{a}_3^\dagger$ & $1+e^{i \varphi}$ & $1-e^{i \varphi}$ & $1-e^{i \varphi}$ &$1+e^{i \varphi}$ \\
        \hline
    \end{tabular}
    \caption{Table representation of the state $\ket{s_\text{out}}$, up to normalization, separating photonic creation operators, acting on the vacuum state $|\overline\emptyset\rangle$ (rows) and spin-qubit states (columns)} 
    \label{tab:statesout}
\end{table}

\subsubsection{Repeat-type patterns}
When photons bunch into a single output mode, single-qubit operations are applied on the spins. These operations can easily be corrected with a local operation to restore the system to its initial state. For instance, if two photons are detected in mode $0$ (first row of Table~\ref{tab:statesout}), the states of the spins is 
\[\alpha \ket{0_{\rm s}0_{\rm s}} - \beta \ket{0_{\rm s}1_{\rm s}} + \gamma \ket{1_{\rm s}0_{\rm s}} - \delta \ket{1_{\rm s}1_{\rm s}},\]
which is equal to $\Z_b \ket{\phi_{ab}}$. Measuring the two photons in mode $0$ thus effectively applies a $\Z$ operation on the second spin qubit. Using the first four rows of Table~\ref{tab:statesout}, one finds that the two-spin states associated with bunching detection patterns are, up to global phase factors:
\begin{align*}
    \left.\begin{matrix}
        (2, 0, 0,0)\\
        (0, 2, 0, 0)
    \end{matrix}\right\} &\longrightarrow \Z_b\ket{\phi_{ab}},\\
    \left.\begin{matrix}
        (0, 0, 2, 0)\\
        (0, 0, 0, 2)
    \end{matrix}\right\} &\longrightarrow \Z_a\ket{\phi_{ab}},\\
\end{align*}
where $(m_0, m_1, m_2, m_3)$ denotes the detection pattern where $m_i$ photons have been detected in the $i^{\rm th}$-mode detector. In both cases, after applying a $\Z$ operation on one of the qubits, one recovers the identity gate.

\subsubsection{Success-type patterns}

When photons are detected in different modes, a non-trivial two-qubit gate is applied on both spins. Denoting $\ket{\Phi^+_{ab}}=\alpha\ket{0_{\rm s}0_{\rm s}}+\delta\ket{1_{\rm s}1_{\rm s}}$ and $\ket{\Psi^+_{ab}}=\beta\ket{0_{\rm s}1_{\rm s}}+\gamma\ket{1_{\rm s}0_{\rm s}}$, the spin states for successful detection patterns are, up to normalization and global phase factors:
\begin{align*}
    \left.\begin{matrix}
        (1, 0, 1, 0)\\
        (0, 1, 0, 1)
    \end{matrix}\right\} &\longrightarrow \cos{\frac{\varphi}{2}}\ket{\Phi^+_{ab}}-i\sin{\frac{\varphi}{2}}\ket{\Psi^+_{ab}},\\
    \left.\begin{matrix}
        (1, 0, 0, 1)\\
        (0, 1, 1, 0)
    \end{matrix}\right\} &\longrightarrow -i\sin{\frac{\varphi}{2}}\ket{\Phi^+_{ab}}+\cos{\frac{\varphi}{2}}\ket{\Psi^+_{ab}}.
\end{align*}
In contrast with the repeat-type detection patterns, a two-qubit operation is applied, determined by the phase parameter $\varphi$. We consider two cases: $\varphi=0$ and $\varphi=\pi/2$.

When $\varphi=0$, the interferometer performs a non-destructive $\Z\Z$ measurement: if the photons are detected in modes 0 and 2 or 1 and 3 (resp.\ 0 and 3 and 1 and 2), the spins are in the state $\ket{\Phi^+_{ab}}$ (resp.\ $\ket{\Psi^+_{ab}}$) corresponding to a $+1$ (resp.\ $-1$) $\Z\Z$-measurement outcome.

When $\varphi=\pi/2$, a $\CZ$ gate is applied after single-qubit corrections. Indeed, up to normalization, the two-spin states are:
\begin{align*}
    \left.\begin{matrix}
        (1, 0, 1, 0)\\
        (0, 1, 0, 1)
    \end{matrix}\right\} &\longrightarrow \ket{\Phi^+_{ab}}-i\ket{\Psi^+_{ab}},\\
    \left.\begin{matrix}
        (1, 0, 0, 1)\\
        (0, 1, 1, 0)
    \end{matrix}\right\} &\longrightarrow -i\ket{\Phi^+_{ab}}+\ket{\Psi^+_{ab}}.
\end{align*}
Applying $\S_a\S_b$ in the first two cases and $\S_a^\dagger\S_b^\dagger$ in the last two, where $\S = \sqrt{\Z}$ is the phase gate, we get:
\begin{align*}
    \left.
    \begin{matrix}
        (1, 0, 1, 0)\\
        (0, 1, 0, 1)
    \end{matrix}
    \right\} 
    & \longrightarrow 
    \begin{matrix*}[l]
        \alpha\ket{0_{\rm s}0_{\rm s}} + \beta\ket{0_{\rm s}1_{\rm s}} \\ 
        \hspace{0.5cm}
        +\gamma\ket{1_{\rm s}0_{\rm s}}-\delta\ket{1_{\rm s}1_{\rm s}},
    \end{matrix*}
    \\
    \left.
    \begin{matrix}
        (1, 0, 0, 1)\\
        (0, 1, 1, 0)
    \end{matrix}
    \right\} 
    & \longrightarrow 
    \begin{matrix*}[l]
        i ( \alpha\ket{0_{\rm s}0_{\rm s}} +\beta\ket{0_{\rm s}1_{\rm s}} \\
        \hspace{0.5cm}
        +\gamma\ket{1_{\rm s}0_{\rm s}}-\delta\ket{1_{\rm s}1_{\rm s}} ) ,
    \end{matrix*}
\end{align*}
which corresponds to a $\CZ$ gate applied on the two spins, up to a global phase factor.

\subsection{Distinguishable photons}
\label{app:distinguishability}

We can generalize the computation of Appendix \ref{app:interferometer} to photons with an additional internal degree of freedom, which corresponds to a witness of distinguishability between photons. The generalized creation operators is $\hat{a}_i^\dagger[\xi_\mathtt{S}]$ and acts on the photonic vacuum as
\begin{equation}
\hat{a}_i^\dagger[\xi_s]|\overline\emptyset\rangle= \ket{1_i}\otimes\ket{\xi_s},    
\end{equation}
where $\ket{1_i}$ corresponds to a single photon in mode $i$ and $\ket{\xi_s}$ models the internal degree of freedom with a label that takes values $s=\{a,b\}$,  indicating which source emitted the photon. Table \ref{tab:statesout_dist} then generalizes Table \ref{tab:statesout}, accounting for labeling.

\begin{table}[h]
\small
    \centering
    \begin{tabular}{|c|c|c|c|c|}
    \hline  
         & $\alpha \ket{0_{\rm s}0_{\rm s}} $&$\beta \ket{0_{\rm s}1_{\rm s}} $ &$ \gamma \ket{1_{\rm s}0_{\rm s}}$ &$\delta \ket{1_{\rm s}1_{\rm s}}$  \\
         \hline 
        $ \hat{a}_0^\dagger[\xi_a] \hat{a}_0^\dagger[\xi_b] $&  $1$ & $-1$ & $1$ & $-1$ \\
        $ \hat{a}_1^\dagger[\xi_a] \hat{a}_1^\dagger[\xi_b] $&  $-1$ & $1$ & $-1$ & $1$ \\
        $ \hat{a}_2^\dagger[\xi_a] \hat{a}_2^\dagger[\xi_b] $&  $e^{i \varphi}$ & $e^{i \varphi}$ & $-e^{i \varphi}$ & $-e^{i \varphi}$ \\
        $ \hat{a}_3^\dagger[\xi_a] \hat{a}_3^\dagger[\xi_b] $&  $-e^{i \varphi}$ & $-e^{i \varphi}$ & $e^{i \varphi}$ & $e^{i \varphi}$ \\
        \hline
        $\hat{a}_0^\dagger[\xi_a] \hat{a}_1^\dagger[\xi_b]$ & $-1$ & $1$ & $-1$ &$1$ \\
        $\hat{a}_0^\dagger[\xi_b] \hat{a}_1^\dagger[\xi_a]$ & $1$ & $-1$ & $1$ &$-1$ \\
        $\hat{a}_2^\dagger[\xi_a] \hat{a}_3^\dagger[\xi_b]$ & $e^{i\varphi}$ & $e^{i\varphi}$ & $-e^{i\varphi}$ &$-e^{i\varphi}$ \\
        $\hat{a}_2^\dagger[\xi_b] \hat{a}_3^\dagger[\xi_a]$ & $-e^{i\varphi}$ & $-e^{i\varphi}$ & $e^{i\varphi}$ &$e^{i\varphi}$ \\
        \hline
        $\hat{a}_0^\dagger[\xi_a] \hat{a}_2^\dagger[\xi_b]$ & $1$ & $1$ & $1$ &$1$ \\
        $\hat{a}_0^\dagger[\xi_b] \hat{a}_2^\dagger[\xi_a]$ & $e^{i \varphi}$ & $-e^{i \varphi}$ & $-e^{i \varphi}$ &$e^{i \varphi}$ \\
        $\hat{a}_0^\dagger[\xi_a] \hat{a}_3^\dagger[\xi_b]$ & $1$ & $1$ & $1$ &$1$ \\
        $\hat{a}_0^\dagger[\xi_b] \hat{a}_3^\dagger[\xi_a]$ & $-e^{i \varphi}$ & $e^{i \varphi}$ & $e^{i \varphi}$ &$-e^{i \varphi}$ \\
        $\hat{a}_1^\dagger[\xi_a] \hat{a}_2^\dagger[\xi_b]$ & $1$ & $1$ & $1$ &$1$ \\
        $\hat{a}_1^\dagger[\xi_b] \hat{a}_2^\dagger[\xi_a]$ & $-e^{i \varphi}$ & $e^{i \varphi}$ & $e^{i \varphi}$ &$-e^{i \varphi}$ \\
        $\hat{a}_1^\dagger[\xi_a] \hat{a}_3^\dagger[\xi_b]$ & $1$ & $1$ & $1$ &$1$ \\
        $\hat{a}_1^\dagger[\xi_b] \hat{a}_3^\dagger[\xi_a]$ & $e^{i \varphi}$ & $-e^{i \varphi}$ & $-e^{i \varphi}$ &$e^{i \varphi}$ \\
        \hline
    \end{tabular}
    \caption{Table representation of the state $\ket{s_\text{out}}$, up to normalization, separating photonic creation operators, acting on the vacuum state $|\overline\emptyset\rangle$ (rows) and spin-qubit states (columns)} 
    \label{tab:statesout_dist}
\end{table}

Let's illustrate with the \suc{} detection pattern $m=(1,0,1,0)$. The term yielding such pattern corresponds to the first two rows of the bottom part of Table \ref{tab:statesout_dist}. The associated state reads
\begin{widetext}
    \begin{equation}
    \begin{aligned}
        \ket{s_{m}}  
    &=\left(\hat{a}_0^\dagger[\xi_a] \hat{a}_2^\dagger[\xi_b]\otimes\Id_a\Id_b
    +e^{i\varphi}\hat{a}_0^\dagger[\xi_b] \hat{a}_2^\dagger[\xi_a]\otimes\Z_a\Z_b\right)\ket{\phi_{ab}} \nonumber \\
    &=\ket{1_0 1_2} \otimes \big( \ket{\xi_a\xi_b}\otimes\ket{\phi_{ab}}+ e^{i\varphi}\ket{\xi_b\xi_a}\otimes\Z_a\Z_b\ket{\phi_{ab}}\big)
    \end{aligned}
\end{equation}
\end{widetext}

The spins state after such a detection event is obtained by tracing out the photonic degree of freedom of the density matrix $\rho_{m}=\op{s_{m}}$. After some rewriting, this reads 

\begin{align*}
    {\rm Tr}_{ph}(\rho_{m})&=\left(1-\frac{D}{2}\right)\mathsf{K}_+(\varphi)\ketbra{\phi_{ab}}{\phi_{ab}}\mathsf{K}_+(\varphi)^\dagger\\
    &\quad+ \frac{D}{2}\mathsf{K}_-(\varphi)\ketbra{\phi_{ab}}{\phi_{ab}}\mathsf{K}_-(\varphi)^\dagger,
\end{align*}
where $K_\pm(\varphi)=(\Id_a\Id_b\pm e^{i\varphi}\Z_a\Z_b)/2$, $D=1-M$, and $M=|\langle \xi_a|\xi_b\rangle|^2$ represents the mean wave-packet overlap between the two photons. When $\varphi=\pi/2$, we get (upon normalization):
\begin{align}
    {\rm Tr}_{ph}(\rho_{m}) = & \left(1-\frac{D}{2}\right)\S_b^\dagger\S_a^\dagger\CZ \rho \CZ\S_a\S_b \\ 
    & + \frac{D}{2}\S_b\S_a\CZ \rho \CZ\S_b^\dagger\S_a^\dagger, \nonumber
\end{align}
which is equivalent to performing the corrected $\CZ$ operation $\mathcal{C}:\rho \mapsto \S_b^\dagger\S_a^\dagger\CZ \rho \CZ\S_a\S_b$, as in the noiseless case, parasitized by a noise channel $\rho \rightarrow (1-D/2)\rho+D\Z_b\Z_a\rho\Z_a\Z_b/2$. Overall, when the detection pattern is $(1,0,1,0)$, the applied channel is thus
\begin{align}
\mathcal{C}_{dist.}^{\CZ}[\rho] = & \left(1-\frac{D}{2}\right) \S_b^\dagger\S_a^\dagger\CZ \rho \CZ\S_a\S_b  \\
& + \frac{D}{2} \Z_b\Z_a\S_b^\dagger\S_a^\dagger\CZ \rho \CZ\S_a\S_b\Z_a\Z_b, \nonumber    
\end{align}
while, when $\varphi=0$, the channel is
\begin{equation}
    \mathcal{C}^{(+)}[\rho] = \left(1 - \frac{D}{2}\right) {\Pi}_{+}\rho{\Pi}_{+} + \frac{D}{2} {\Pi}_{-}\rho{\Pi}_{-}, 
\end{equation}
where ${\Pi}_{\pm}=(\Id_a\Id_b\pm\Z_a\Z_b) / 2$. This channel corresponds to a measurement error with probability $D/2$. Indeed, in the fully indistinguishable case, the pattern $(1, 0, 1, 0)$ corresponds to projecting solely on the $+1$ eigenvalue of $\Z_a\Z_b$. In the distinguishable case, we would still measure $(1, 0, 1, 0)$ and thus record $+1$ as the measurement of $\Z_a\Z_b$, but there is a possibility for the spins to be projected on the $-1$ eigenvalue instead. Note that, for $\varphi=0$, the same channel applies for the $(0,1,0,1)$ pattern, but that, for the $(1,0,0,1)$ and $(0,1,1,0)$ patterns, the channel is:
\begin{equation}
    \mathcal{C}^{(-)}[\rho] = \left(1 - \frac{D}{2}\right) {\Pi}_{-}\rho{\Pi}_{-} + \frac{D}{2} {\Pi}_{+}\rho{\Pi}_{+}.
\end{equation}

\subsection{Photon loss}
\label{app:loss}
As discussed in the main text, \lossone\ and \losstwo\ patterns correspond to losing respectively one or two photons in the system. When one photon is lost, the channel 
\begin{equation}
  \mathcal{C}_{loss}^{(1)}[\rho] = \frac{\mathcal{C}_a[\rho] + \mathcal{C}_b[\rho]}{2} \label{eq:ap_c_rus1Photon}
\end{equation} is applied (with corrections depending on the specific outcome pattern) while when both photons are lost, \begin{equation}
    \mathcal{C}_{loss}^{(\infty)}[\rho] = \mathcal{C}_a \circ \mathcal{C}_b[\rho].
    \label{eq:ap_c_rusNoPhoton}
\end{equation}
is applied. Due to the possibility of getting \bunch{} patterns, the RUS subroutine may be applied several times. If after several repetitions of the subroutine at least one two-photon loss event occurred, the total noise channel  $\mathcal{C}_{loss}^{(\infty)}$ is applied regardless of the number of one-photon loss events since  $\mathcal{C}_{loss}^{(1)} \circ \mathcal{C}_{loss}^{(\infty)} = \mathcal{C}_{loss}^{(\infty)} \circ \mathcal{C}_{loss}^{(1)} = \mathcal{C}_{loss}^{(\infty)}$. This can be straightforwardly understood as $\mathcal{C}_{loss}^{(1)}$ corresponds to a phase erasure of one of the two qubits involved in the channel while $\mathcal{C}_{loss}^{(\infty)}$ corresponds to the phase erasure of the two qubits. If at most one photon is lost per repetition (no two-photon loss event), the channel applied on the spin qubits is $\mathcal{C}_{loss}^{(k)}$ where $k$ is the total number of photons lost and $\left(\cdot\right)^{(k)}$ denotes the $k$-fold composition. Every time a single-photon loss event occurs, the phase of one of the two spin qubits is erased. After $k$ single -photon loss events, there is a probability $p = 1/2^k$ that we lose the phase of only qubit $a$ (respectively qubit $b$) but in the remaining case, we lose both phases. Therefore,
\begin{align}
    \mathcal{C}_{loss}^{(k)} &= \frac{1}{2^{k-1}} \frac{\mathcal{C}_a + \mathcal{C}_b}{2} + \left(1 - \frac{1}{2^{k-1}}\right) \mathcal C_a \circ \mathcal C_b\\
\intertext{which reads when applied to a density matrix $\rho$:}
    \mathcal{C}_{loss}^{(k)}[\rho] &= \frac{1}{4}\left(1 + \frac{1}{2^{k-1}}\right)\rho \nonumber\\& \quad+ \frac{1}{4}\left(1 - \frac{1}{2^{k-1}}\right) \Z_a\Z_b\rho\Z_a\Z_b  \label{eq:ap_c_ruskPhoton}\\
    &\quad  + \frac{1}{4}\left(\Z_a \rho \Z_a + \Z_b \rho \Z_b\right).\nonumber
\end{align}
Note that $\mathcal{C}_{loss}^{(k)}\to_{k\to\infty}\mathcal{C}_{loss}^{(\infty)}$ which is expected since as we get more single-photon loss events, the chances of erasing both phases increase. 

By definition, a $\Z\Z$-measurement  projects the state in a subspace stabilized by $\pm \Z_a\Z_b$ (with $\pm$ depending on the measurement outcome). Notice that the first two terms of Eq. \eqref{eq:ap_c_ruskPhoton} act trivially on the state after $\MZZ$, hence only the last two terms can lead to an error. In the case of a RUS $\MZZ$, the error channel of Eq. \eqref{eq:ap_c_ruskPhoton}, is thus equivalent to 
\begin{equation}
    \frac{1}{4} (2\rho + \Z_a \rho \Z_a + \Z_b \rho Z_b) =  \mathcal{C}_{loss}^{(1)}[\rho]. 
\end{equation}
Since in that case both error channels $\mathcal{C}_{loss}^{(k)}$ and $\mathcal{C}_{loss}^{(\infty)}$ act in the same way, we can treat them jointly and only define a total $p_{\rm RUS}$ failure probability of having strictly less than two photons detected. This discussion only applies to the RUS $\MZZ$ measurement, not to the RUS $\CZ$ gate.

\section{Simulation of the RUS gate}
\label{sec:simu_rus_gate}
This appendix is dedicated to a more formal description of the RUS gate and the computation of the logical error rates of QEC protocols where entanglement is generated using RUS subroutines.

\subsection{Mathematical description}

In the following, we propose a description of a noisy RUS gate in terms of heralded channels. 
Let us consider the following encoding channel:
\begin{equation}
\begin{aligned}
    \mathcal{C}_{enc}:\mathcal{M}_{4,4}(\mathbb{C})&\rightarrow\mathcal{M}_{4,4}(\mathbb{C})\otimes\mathcal{M}_{4,4}(\mathbb{C})\\
\rho&\mapsto\rho\otimes\ketbra{00}{00}.
\end{aligned}
\end{equation}
This channel appends two qubits to the density matrix. 
These ``qubits'' are fictitious  and will act as ``repeat" flags and measurement records in the RUS cycle. They represent the classical register on which we will store the classical information corresponding to the measurement outcomes and the operations that we should perform on them should be the one allowed solely by Boolean algebra and not quantum mechanics.

\begin{table*}[ht!]
    \centering
    \begin{tblr}{|c||c|c|c|c|c|c|}
        \hline
          & \SetCell[c=4]{} $\mathtt{R}$ & & & &\SetCell[c=2]{} $\mathtt{Success}$ \\
         \hline
         Type & \SetCell[c=2]{} \bunch &  & \lossone & \losstwo & \sucp & \sucm \\
         \hline
         \SetCell[r=4]{} Pattern & (2, 0, 0, 0) &   & (1, 0, 0, 0) & \SetCell[r=4]{}(0, 0, 0, 0)                &  & \\
        & (0, 2, 0, 0) & (1, 1, 0, 0) & (0, 1, 0, 0) &              &  (1, 0, 1, 0) & (1, 0, 0, 1)\\
        & (0, 0, 2, 0) &     (0, 0, 1, 1)         & (0, 0, 1, 0) &              &   (0, 1, 0, 1)           &  (0, 1, 1, 0)           \\
                 & (0, 0, 0, 2) &              & (0, 0, 0, 1) &              &              &             \\
         \hline
         Probability & $(1-\varepsilon)^2\dfrac{2-D}{16}$ & $(1-\varepsilon)^2\dfrac{D}{8}$ & $\dfrac{\varepsilon(1-\varepsilon)}{2}$ & $\varepsilon^2$ & $\dfrac{(1-\varepsilon)^2}{4}p_+$ & $\dfrac{(1-\varepsilon)^2}{4}p_-$\\
         \hline 
         Channels & $\mathcal{I}$ & $\mathcal{I}$ & $\mathcal{C}_{loss}^{(1)}$ & $ \mathcal{C}_{loss}^{(\infty)}$ & $\mathcal{C}^{(+)}$ & $\mathcal{C}^{(-)}$\\
         \hline
    \end{tblr}
    \caption{Possible detection patterns for a RUS gate and their associated channel on the density matrix of the spins for the noisy $\MZZ$ RUS gate. The different types corresponds to sets of measurement patterns, e.g.\ $(2, 0, 0, 0) \in \bunch$. $\mathcal{I}$ is the identity map. The probabilities $p_+=|\alpha|^2+|\delta|^2$ and $p_-=|\beta|^2+|\gamma|^2$ correspond to the probabilities of measuring operator $+\Z\Z$ or $-\Z\Z$ for a pure state $\ket{\phi_{ab}}=\alpha\ket{0_{\rm s}0_{\rm s}}+\beta\ket{0_{\rm s}1_{\rm s}}+\gamma\ket{1_{\rm s}0_{\rm s}}+\delta\ket{1_{\rm s}1_{\rm s}}$.}
    \label{tab:herald_channels}
\end{table*}

All the heralded channels for the RUS $\MZZ$ are displayed in Table~\ref{tab:herald_channels}. The potential unitary corrections are included in the channels. All these channels are composed either of $\Id$ or $\Z$ operators and thus all commute. Errors induced by decoherence are also $\Z$-type errors that depend on the total time of a RUS gate and commute with all these channels as well.

In the following, we focus on the $\MZZ$ gate, but the $\CZ$ RUS gate can be described analogously. The only difference is that the associated channel to $\mathtt{S} = \mathtt{Success}_+ \cup \mathtt{Success}_-$ is $\mathcal C_{dist.}^{\CZ}$ and there is no measurement outcome (therefore no need for a second fictitious qubit).

For the $\mathsf{M_{ZZ}}$ RUS gate, we will group these patterns into two categories: 
\begin{itemize}
    \item the ``repeat'' set $\mathtt{R}$ corresponding to all the patterns going from step (3) to step (1) in Fig.~\ref{fig:RUS}(b), i.e. patterns from \bunch, \lossone\ and \losstwo;
    \item  and the $\mathtt{Success}$ set corresponding to all the patterns going from step (3) to noisy $\MZZ$ in Fig.~\ref{fig:RUS}(b), partitioned in \sucp $\cup$ \sucm.
\end{itemize}

These channels are conditioned by a measurement pattern $m$ and we will call a generic channel $\mathcal{C}_m$. For example, $\mathcal{C}_{(0,1,0,0)}=\mathcal{C}_{loss}^{(1)}$. Let us describe the ``RUS cycle channel'' as acting onto a previous encoded density matrix through the map $\mathcal{C}^{(1)}_{\rm cycle}$. We start by describing its action on $\rho\otimes\ketbra{00}{00}$: 
$$
\begin{aligned}
\mathcal{C}^{(1)}_{\rm cycle}[\rho\otimes\ket{00}\bra{00}]&=\sum_{m\in R}\mathcal{C}_m[\rho]\otimes\ketbra{00}{00}\\
&+\sum_{m\in S}\mathcal{C}_m[\rho]\otimes\ketbra{1d(m)}{1d(m)}.\\
\end{aligned}$$
Here $d: m \in S \mapsto d(m)=\delta_{(m \in \mathtt{success_-})} \in \mathbb F_2$ is the decision map (with $\delta_A$ equal to $1$ if $A$ is true and $0$ otherwise).
It clarifies the information encoded on the fictitious qubits: the first one encodes whether or not the measurement pattern corresponds to a successful outcome. If it is in the $0$ state, we should continue doing RUS gates, if it is in the $1$ state, we should stop. The second fictitious qubit corresponds to the measurement outcome bit value.
In this discussion, we decided for simplicity not to consider the more accurate description of a classical register which keeps track of which  patterns in the set $\mathtt{R}$ is applied, thus losing information about the precise error channel. 
We also want to consider multiple trials, described by the channel $\mathcal C^{(k)}_{\rm cycle} = \mathcal C^{(1)}_{\rm cycle} \circ \cdots \circ\mathcal C^{(1)}_{\rm cycle}$ (composed $k$ times). Since we do not need to repeat the measurement after a success pattern $m \in S$, corresponding to the first qubit value $1$, we need:
\begin{equation*}
    \mathcal{C}^{(1)}_{\rm cycle}(\rho\otimes\ketbra{1 m}{1 m})=\rho\otimes\ketbra{1 m}{1 m}.
\end{equation*}
These are the only type of input state that $\mathcal{C}^{(1)}_{\rm cycle}$ will see during a RUS gate, and the rest of the terms can be taken arbitrarily.

We can see by induction that composing the previous map $k$ times leads to 
\begin{widetext}
    \begin{equation*}
\mathcal{C}^{(k)}_{\rm cycle}[\rho\otimes\ketbra{00}{00}] = \mathcal{C}_\mathtt{R}^{(k)}[\rho]\otimes \ketbra{00}{00} +\sum_{m\in \mathtt{S}}\mathcal{C}_m\circ\left(\sum_{j=0}^{k-1}\mathcal{C}_\mathtt{R}^{(j)}\right)[\rho]\otimes \ketbra{1d(m)}{1d(m)},
\label{eq_c_rus_k}
\end{equation*}
\end{widetext}
where $\mathcal{C}_\mathtt{R}=\sum_{m\in \mathtt{R}}\mathcal{C}_m$ is the aggregated channel of the $\mathtt{R}$ types detection patterns. 

\begin{table*}[t!]
    \centering
    \begin{tblr}{vline{1,3,4,5}={1pt}, vline{2}={1.5pt}, cells={c, m}, row{1-7}={1cm}}
        \hline
         Event & Gate type & Probability & Validity \\
         \hline
         \SetCell[r=2]{} $(N=k)\cap (T=t)$ & $\CZ$ & \SetCell[r=2]{} $\displaystyle p_s\binom{t-1}{k}p_{1}^k p_r^{t-1-k}$ & \SetCell[r=2]{} \parbox{2cm}{$\begin{aligned}
                        0\leq k \leq t-1 \\      
                        1\leq t\leq t_{\max}
                    \end{aligned}$}\\
         \hline
         & $\MZZ$ \\
         \hline
        \SetCell[r=2]{} $(N=\overline{\infty})\cap (T=t)$ & $\CZ$ &  $\displaystyle p_2(p_r+p_{1})^{t-1}$ & $1\leq t\leq t_{\max}$\\
         \hline
         & $\MZZ$ & $\displaystyle p_s\sum_{j=1}^{t-1}\binom{t-1}{j}p_{2}^j (p_r+p_{1})^{t-1-j}$ & $1<t\leq t_{\max}$\\
         \hline
         \SetCell[r=2]{} $(N=a)\cap (T=t)$ & $\CZ$ &  $\displaystyle \delta\left(t-t_{\max}\right)(1-p_s-p_2)^{t_{\max}}$ & \SetCell[r=2]{} $1<t\leq t_{\max}$\\
         \hline
         & $\MZZ$ & $\displaystyle \delta\left(t-t_{\max}\right)(1-p_s)^{t_{\max}}$ \\
         \hline
    \end{tblr}
    \caption{The probabilities associated to the different events of the joint distribution of $(N,T)$. The probabilities $p_r=(1-\varepsilon)^2/2$, $p_s=(1-\varepsilon)^2/2$, $p_1=2\varepsilon(1-\varepsilon)$, and $p_2=\varepsilon^2$ correspond respectively to the probabilities to observe a \bunch, \suc, \lossone, and \losstwo\ event. It is straightforward to check that the probabilities of the distribution of $(N,T)$ sum to one, taking the range of validity (last column) into account.}
    \label{tab:joint_distributions}
\end{table*}

\subsubsection{Infinitely many RUS trials}
If one allows for an infinite amount of trials, then the RUS channel $\mathcal{C}_{RUS} = \mathcal{C}^{(\infty)}_{\rm cycle}$ can be described by 
\begin{equation}
\begin{aligned}
    \mathcal{C}_{RUS}[\rho\otimes\ket{00}\bra{00}]&=\sum_{m\in S}\mathcal{C}_m\circ(\mathcal{I}-\mathcal{C}_\mathtt{R})^{(-1)}[\rho]\\ &\qquad\otimes\ketbra{1d(m)}{1d(m)},
\end{aligned}
\end{equation}
Indeed, because $\mathcal{C}_\mathtt{R}^{(k)}$ is not trace-preserving, $\mathcal{C}_\mathtt{R}^{(k)}\rightarrow_{k\to\infty}0$ and $\sum_{j=0}^{k-1}\mathcal{C}_\mathtt{R}^{(j)}$ converges towards $(\mathcal{I}-\mathcal{C}_\mathtt{R})^{(-1)}$ \footnote{Note here that $(\mathcal{I}-\mathcal{C}_\mathtt{R})^{(-1)}$ is purely a notation that should be understood as $\sum_{j=0}^{\infty}\mathcal{C}_\mathtt{R}^{(j)}$}. It is convenient to observe that (a)  $\mathcal{C}_\mathtt{R}^{(k)}$ commutes with any of the $\mathcal{C}_m$ for $m\in S$ and $(b)$ $\mathrm{Tr}((\mathcal{I}-\mathcal{C}_\mathtt{R})^{(-1)})=(1-p_\mathtt{R})^{-1}=p_\mathtt{S}^{-1}$ (where $p_\mathtt{R}$ and $p_\mathtt{S}$ are the aggregated probability of $\mathtt{R}$ and $\mathtt{S}$ type detection events) such that
$\mathcal{C}_{RUS}=p_\mathtt{S}(\mathcal{I}-\mathcal{C}_\mathtt{R})^{(-1)}\circ\mathcal{C}_{RUS}^{(\mathtt{S})}$.  with
\begin{equation*}
    \mathcal{C}^{(\mathtt{S})}_{RUS}(\rho\otimes\ketbra{00}{00})=\frac{1}{p_\mathtt{S}}\sum_{m\in S}\mathcal{C}_m\otimes\ketbra{d(m)}{d(m)}.
\end{equation*}
By expanding the channels $\mathcal C_m$ for $m\in\mathtt{S}$, we can write it as
\begin{align*}
    \mathcal{C}^{(\mathtt{S})}_{RUS}(\rho\otimes\ket{00}\bra{00})&=\frac{1}{p_\mathtt{S}}\Pi_+ \rho \Pi_+\otimes\mathcal{C}_{\X,D}\left[\ketbra{0}{0}\right]
\\&+\frac{1}{p_\mathtt{S}}\Pi_- \rho \Pi_-\otimes\mathcal{C}_{\X,D}\left[\ketbra{1}{1}\right],
\end{align*}
with $\mathcal{C}_{\X,D}[\rho]=(1-D/2)\rho+D/2\,\X\rho\X$ a bit flip channel with probability $D/2$. The channel $\mathcal{C}^{(\mathtt{S})}_{RUS}$ is thus exactly an $\MZZ$ channel with a measurement error probability $D/2$. The channel $(\mathcal I - \mathcal C_\mathtt{R})^{(-1)}$ only depends on the photon loss rate $\varepsilon$, and corresponds to the application of the error channels induced by photon loss and bunching patterns.

\subsubsection{Finite amount of trials}
In the more realistic case, we cannot consider infinitely many trials, therefore it is possible to end a RUS gate without having a measurement outcome: $\mathcal{C}_\mathtt{R}^{(k)}$ in Eq.~\eqref{eq_c_rus_k} cannot be considered equal to zero. This is the \abo\ event from the main text.

\subsection{Sampling the error channels and computing the logical error rate}
\label{ap:simulation}

To compute the logical error rate, we need to simulate QEC circuits where entangling gates are replaced by their RUS counterparts. In the following, we explain how we can efficiently simulate the RUS error channel and how we use it to estimate the logical error rate of the QEC protocols.

\subsubsection{Equivalent description of the RUS channel}

As explained in the main text, we can describe the loss channel with two random variables:
\begin{enumerate}
    \item A positive random variable $N\in\mathbb{N}$ that records the number of single-photon loss events throughout the RUS execution. We can extend the domain of $N$ to $\mathbb{N}\cup\{\overline{\infty}\}$ where the event $(N=\overline{\infty})$ corresponds to having observed a two-photon loss event at least once throughout the process. Finally, we extend again the domain of $N$ with an additional ``abort'' event $(N=a)$. This event encodes the failure of the RUS gate to reach a stopping outcome after $t_{\max}$ allowed trials.
    \item A strictly positive random variable $T$ capped by a maximum number of allowed trials $t_{\max}$ that records the number of trials required before stopping the RUS cycle.
\end{enumerate}
These two random variables are not independent since for instance $(N=a)\Rightarrow(T=t_{\max})$. Therefore, simulating a RUS gate loss channel amounts to sampling from the joint distribution of $(N,T)$, see Table~\ref{tab:joint_distributions} for the full distribution. The decoherence channel is applied after a round of measurements or a round of entangling CZ gates. If there are $L$ operations (measurement or entangling gates) in the round, the round time $\tau$ will be a random variable given by
\[\tau = \max_{1\leq i\leq L}\left(T_i\right)\cdot t_{cycle},\]
with $T_i$ the number of RUS trials for operation $i$. This round time will be used to compute the decoherence-induced error channel. Finally, distinguishability-induced errors are added to the system whenever the associated gate has succeeded during the execution.






\subsubsection{Estimation of the logical error rate}

The logical error rate $\varepsilon_L$ is defined as the probability not to retrieve the correct logical state after the recovery operation prescribed by syndrome decoding. This quantity is typically estimated using a standard sample mean estimator obtained by $(a)$ randomly sampling the errors of a QEC circuit multiple times; $(b)$ for each sample $i$, recording the syndrome $\vec{m}_i$ and the output logical state $l_i$; $(c)$ decoding the syndrome, thus obtaining a decoder-guessed logical state $d_i$; $(d)$ averaging the observed mismatch $d_i\oplus l_i$ ($\oplus$ is the standard modulo-2 addition).

In our case, the situation is slightly more convoluted, since the error channels that should be applied after each entangling operation are themselves random, because of the RUS subroutines. To circumvent this issue, we generate random instance of the QEC circuit where each error channel is fixed. The random instances are generated by sampling $N_g$ (with $N_g$ the number of entangling gates in the circuit) times the pair $(N,T)$. For each random circuit instance $c_i$ we estimate a $c_i$-specific logical error rate $\hat{\varepsilon}_L(c_i)$ with a sample mean estimator following the above steps
\begin{equation}
    \hat{\varepsilon}_L(c_i)=\frac{1}{N_S}\sum_{j=1}^{N_\mathtt{S}}d_{ij}\oplus l_{ij},
\end{equation}
where $N_S$ is the number of syndrome generated, $d_{ij}$ (resp.\ $l_{ij}$) the decoder outcome (resp.\ the logical output) for syndrome $1\leq i\leq N_S$ and circuit $c_j$. The overall logical error rate $\hat{\varepsilon}_L$ is estimated by averaging $\hat{\varepsilon}_L(c_i)$ over all the generated circuits
\begin{equation}
    \hat{\varepsilon}_L=\frac{1}{N_c}\sum_{i=1}^{N_c}\hat{\varepsilon}_L(c_i),
\end{equation}
with $N_c$ the number of generated circuits. Fig. \ref{fig:ler_estimation} describes the overall process.

\begin{figure*}
\centering
\small
\begin{tikzpicture}
    \node[scale=0.7] (cinit) at (-2,0) {
            \begin{quantikz}
                &  & \gate[2, style={fill=gray!20}, label style={black, rotate=90, label={[draw, very thick, circle, inner sep=2pt, xshift=20pt, yshift=30pt]:{3}}}]{\text{RUS}}  & \\
                & \gate[2, style={fill=gray!20}, label style={black, rotate=90, label={[draw, very thick, circle, inner sep=2pt, xshift=20pt, yshift=30pt]:{1}}}]{\text{RUS}} &   & \meter{m_1}\\
                &  &  & \\[0.2cm]
                & \gate[2, style={fill=gray!20}, label style={black, rotate=90, label={[draw, very thick, circle, inner sep=2pt, xshift=20pt, yshift=30pt]:{2}}}]{\text{RUS}} &  &\\
                &  & \gate[2, style={fill=gray!20}, label style={black, rotate=90, label={[draw, very thick, circle, inner sep=2pt, xshift=20pt, yshift=30pt]:{4}}}]{\text{RUS}} & \meter{m_2}\\
                &  &  &
            \end{quantikz}
        };

    \node[scale=0.7] (c1) at (3.5,7) {
            \begin{quantikz}
                &  & & \ctrl{1}  & \gate[2, style={fill=ForestGreen!20}]{\mathcal{C}_{dist.}} &  \gate[1, style={fill=NavyBlue!20}]{\mathcal{C}_{deco.}^{(\tau_1)}}&\\
                & \ctrl{1} & \gate[2, style={fill=ForestGreen!20}]{\mathcal{C}_{dist.}} & \control{} & & \gate[1, style={fill=NavyBlue!20}]{\mathcal{C}_{deco.}^{(\tau_1)}}&\meter{m_1}\\
                & \control{} & & & &\gate[1, style={fill=NavyBlue!20}]{\mathcal{C}_{deco.}^{(\tau_1)}}&\\
                & \ctrl{1} & \gate[2, style={fill=ForestGreen!20}]{\mathcal{C}_{dist.}} & & & \gate[1, style={fill=NavyBlue!20}]{\mathcal{C}_{deco.}^{(\tau_1)}}&\\
                & \control{} & & \ctrl{1} & \gate[2, style={fill=ForestGreen!20}]{\mathcal{C}_{dist.}} &\gate[1, style={fill=NavyBlue!20}]{\mathcal{C}_{deco.}^{(\tau_1)}} &\meter{m_2}\\
                &  & & \control{} & &\gate[1, style={fill=NavyBlue!20}]{\mathcal{C}_{deco.}^{(\tau_1)}}&
            \end{quantikz}
        };

    \node[ultra thick] (dots1) at (3.5,3.5) {\bf $\mathbf{\vdots}$};

    \node[scale=0.7] (cj) at (3.5,0) {
            \begin{quantikz}
                &  &  &  \ctrl{1}  & \gate[2, style={fill=ForestGreen!20}]{\mathcal{C}_{dist.}} &\gate[1, style={fill=NavyBlue!20}]{\mathcal{C}_{deco.}^{(\tau_j)}} &\\
                & \ctrl{1} & \gate[2, style={fill=ForestGreen!20}]{\mathcal{C}_{dist.}} & \control{}  & &\gate[1, style={fill=NavyBlue!20}]{\mathcal{C}_{deco.}^{(\tau_j)}} &\meter{m_1}\\
                & \control{} && & & \gate[1, style={fill=NavyBlue!20}]{\mathcal{C}_{deco.}^{(\tau_j)}}&\\
                & \ctrl{1} & \gate[2, style={fill=ForestGreen!20}]{\mathcal{C}_{dist.}}& \gate[2, style={fill=BrickRed!20}]{\mathcal{C}_{loss}^{(1)}} &&\gate[1, style={fill=NavyBlue!20}]{\mathcal{C}_{deco.}^{(\tau_j)}}&\\
                & \control{} && & \gate[2, style={fill=BrickRed!20}]{\mathcal{C}_{abor.}} &\gate[1, style={fill=NavyBlue!20}]{\mathcal{C}_{deco.}^{(\tau_j)}} &\meter{m_2}\\
                &  && &  &\gate[1, style={fill=NavyBlue!20}]{\mathcal{C}_{deco.}^{(\tau_j)}}&
            \end{quantikz}
        };

    \node[ultra thick] (dots2) at (3.5,-3.25) {\bf $\mathbf{\vdots}$};

    \node[scale=0.7] (cn) at (3.5,-7) {
            \begin{quantikz}
                &  & & & \ctrl{1}  & \gate[2, style={fill=BrickRed!20}]{\mathcal{C}_{loss}^{(\infty)}} &\gate[1, style={fill=NavyBlue!20}]{\mathcal{C}_{deco.}^{(\tau_{n_c})}} &\\
                & \ctrl{1} & \gate[2, style={fill=ForestGreen!20}]{\mathcal{C}_{dist.}}& \gate[2, style={fill=BrickRed!20}]{\mathcal{C}_{loss}^{(2)}} & \control{}  & & \gate[1, style={fill=NavyBlue!20}]{\mathcal{C}_{deco.}^{(\tau_{n_c})}}&\meter{m_1}\\
                & \control{} & & & & & \gate[1, style={fill=NavyBlue!20}]{\mathcal{C}_{deco.}^{(\tau_{n_c})}} & \\
                & \ctrl{1} & \gate[2, style={fill=ForestGreen!20}]{\mathcal{C}_{dist.}} & & & & \gate[1, style={fill=NavyBlue!20}]{\mathcal{C}_{deco.}^{(\tau_{n_c})}} & \\
                & \control{} & & & \ctrl{1} & \gate[2, style={fill=ForestGreen!20}]{\mathcal{C}_{dist.}}& \gate[1, style={fill=NavyBlue!20}]{\mathcal{C}_{deco.}^{(\tau_{n_c})}} & \meter{m_2} \\
                & & & & \control{} & & \gate[1, style={fill=NavyBlue!20}]{\mathcal{C}_{deco.}^{(\tau_{n_c})}} &
            \end{quantikz}
        };

    \node[scale=0.8, align=center, inner sep=0pt, minimum width=1.2cm] (s1) at (7.8,7) {\begin{tabular}{c c}
         $m_1$ & $m_2$ \\\hline
         0 & 1 \\
         1 & 1 \\
         $\vdots$ & $\vdots$ \\
         0 & 0 \\
         0 & 1 \\
         1 & 1 \\
         $\vdots$ & $\vdots$ \\
         1 & 0 \\
         0 & 1
    \end{tabular}};
    
    \node[scale=0.8, align=center, inner sep=0pt, ] (d1) at (9.5, 7) {\begin{tabular}{c}
         $d\oplus l$ \\\hline
         0 \\
         1 \\
         $\vdots$ \\
         0\\
         0 \\
         1 \\
         $\vdots$ \\
         1 \\
         0
    \end{tabular}};

    \node[scale=0.8, align=center, inner sep=0pt, minimum width=1.2cm] (sj) at (7.8, 0) {\begin{tabular}{c c}
         $m_1$ & $m_2$ \\\hline
         1 & 1 \\
         0 & 1 \\
         $\vdots$ & $\vdots$ \\
         1 & 1 \\
         1 & 0 \\
         1 & 1 \\
         $\vdots$ & $\vdots$ \\
         1 & 0 \\
         1 & 1
    \end{tabular}};
    
    \node[scale=0.8, align=center, inner sep=0pt, ] (dj) at (9.5, 0) {\begin{tabular}{c}
         $d\oplus l$ \\\hline
         1 \\
         1 \\
         $\vdots$ \\
         1\\
         0 \\
         1 \\
         $\vdots$ \\
         1 \\
         0
    \end{tabular}};

    \node[scale=0.8, align=center, inner sep=0pt, minimum width=1.2cm] (sn) at (7.8,-7) {\begin{tabular}{c c}
         $m_1$ & $m_2$ \\\hline
         0 & 0 \\
         0 & 1 \\
         $\vdots$ & $\vdots$ \\
         1 & 0 \\
         1 & 0 \\
         0 & 1 \\
         $\vdots$ & $\vdots$ \\
         1 & 0 \\
         0 & 1
    \end{tabular}};
    
    \node[scale=0.8, align=center, inner sep=0pt, ] (dn) at (9.5, -7) {\begin{tabular}{c}
         $d\oplus l$ \\\hline
         0 \\
         1 \\
         $\vdots$ \\
         0\\
         0 \\
         1 \\
         $\vdots$ \\
         1 \\
         1
    \end{tabular}};
    
    \node[scale=1, align=center, inner sep=0pt] (e1) at (11.3, 7) {$\hat{\varepsilon}_L(c_{1})$};
    \node[scale=1, align=center, inner sep=0pt] (ej) at (11.3, 0) {$\hat{\varepsilon}_L(c_{j})$};
    \node[scale=1, align=center, inner sep=0pt] (en) at (11.3, -7) {$\hat{\varepsilon}_L(c_{n_c})$};

    \node[scale=1.1, align=center, inner sep=2pt] (el) at (13.2,0) {$\hat{\varepsilon}_L$};

    \begin{scope}[on background layer]
        \node[draw, very thick, rounded corners=0.5cm, fit=(cn), fill=MidnightBlue!5] (cnn) {};
        \node[draw, very thick, rounded corners=0.5cm, fit=(c1), fill=MidnightBlue!5] (cn1) {};
        \node[draw, very thick, rounded corners=0.5cm, fit=(cj), fill=MidnightBlue!5] (cnj) {};
        \node[draw, very thick, rounded corners=0.5cm, fit=(cinit), fill=Yellow!5] (cninit) {};

        \node[draw, very thick, rounded corners=0.2cm, inner sep=1pt, fit=(s1), fill=Rhodamine!5] (s1n) {};
        \node[draw, very thick, rounded corners=0.2cm, fit=(d1), fill=JungleGreen!5] (d1n) {};
        \node[draw, very thick, rounded corners=0.2cm, inner sep=1pt, fit=(sj), fill=Rhodamine!5] (sjn) {};
        \node[draw, very thick, rounded corners=0.2cm, fit=(dj), fill=JungleGreen!5] (djn) {};
        \node[draw, very thick, rounded corners=0.2cm, inner sep=1pt, fit=(sn), fill=Rhodamine!5] (snn) {};
        \node[draw, very thick, rounded corners=0.2cm, fit=(dn), fill=JungleGreen!5] (dnn) {};

        \node[draw, very thick, rounded corners=0.2cm, fit=(en), fill=BrickRed!5] (enn) {};
        \node[draw, very thick, rounded corners=0.2cm, fit=(ej), fill=BrickRed!5] (ejn) {};
        \node[draw, very thick, rounded corners=0.2cm, fit=(e1), fill=BrickRed!5] (e1n) {};
        \node[draw, very thick, rounded corners=0.2cm, fit=(el), fill=BrickRed!20] (eln) {};
    \end{scope}

    \draw[line width=2pt, my triangle] (cninit) to ($(cninit)+(0,7)$) to (cn1);
    \draw[line width=2pt, my triangle] (cninit) to (cnj);
    \draw[line width=2pt, my triangle] (cninit) to ($(cninit)+(0,-7)$) to (cnn);

    \draw[line width=2pt, my triangle] (cn1) -- (s1n);
    \draw[line width=2pt, my triangle] (s1n) -- (d1n); 
    \draw[line width=2pt, my triangle] (d1n) to node[thin, draw, circle, scale=0.7, pos=0.45, fill=white] (avglab) {$\mu$} (e1n); 

    \draw[line width=2pt, my triangle] (cnj) -- (sjn);
    \draw[line width=2pt, my triangle] (sjn) -- (djn);
    \draw[line width=2pt, my triangle] (djn) to node[thin, draw, circle, scale=0.7, pos=0.45, fill=white]  {$\mu$} (ejn); 

    \draw[->] ($(avglab)+(0.5, -3)$) node[draw, rounded corners=0.2cm, align=center, fill=Gray!5] {Average input\\values}  to (avglab);

    \draw[->] ($(dn)+(-1.5,3)$) node[draw, rounded corners=0.2cm, align=center, fill=Gray!5] {Decoder\\outcome}  to ($(dn)+(-0.3,2.1)$);

    \draw[->] ($(dn)+(1.5,3)$) node[draw, rounded corners=0.2cm, align=center, fill=Gray!5] {Logical\\outcome}  to ($(dn)+(0.3,2.1)$);
    
    \draw[line width=2pt, my triangle] (cnn) -- (snn);
    \draw[line width=2pt, my triangle] (snn) -- (dnn); 
    \draw[line width=2pt, my triangle] (dnn) to node[thin, draw, circle, scale=0.7, pos=0.45, fill=white] {$\mu$} (enn);

    \node[draw, circle, scale=0.8] (avg) at ($(eln)+(-0.85,0)$) {$\mu$};
    \draw[line width=2pt] (e1n) to ($(avg)+(0,7)$) to (avg);
    \draw[line width=2pt] (enn) to ($(avg)+(0,-7)$) to (avg);
    \draw[line width=2pt, my triangle] (ejn) to (avg) to (eln);

    \node[draw, rounded corners=4pt, align=center] at (12.75, 10.2) {Overall\\LER};
    \node[draw, rounded corners=4pt, align=center] at (11.25, 10.2) {Random\\circuit\\LER};
    \node[draw, rounded corners=4pt, align=center] at (9.6, 10.2) {Syndrome\\decoding};
    \node[draw, rounded corners=4pt, align=center] at (7.8, 10.2) {Syndrome\\sampling};
    \node[draw, rounded corners=4pt, align=center] at (3.55, 10.2) {Random circuit creation};
    \node[draw, rounded corners=4pt, align=center] at (-2, 10.2) {Abstract circuit\\definition};

\end{tikzpicture}

\caption{Computation of the logical error rate (LER) estimator $\hat{\varepsilon}_L$ for an example of circuit with 4 $\CZ$ RUS gates. From the abstract, error-free QEC circuit, we sample $L=4$ times the pair of random variables $(N,T)$. Depending on the $i$-th sample of $N$, we attach the corresponding error channel to the $i$-th RUS gate. The overall decoherence depends on the maximum value of the samples of $T$. For each generated circuit $c_j$, we sample $N_S$ syndromes (which correspond to sampling $m_1$ and $m_2$ in this example). We decode each sample and compare the decoder outcome $d$ to the logical outcome $l$ of the circuit. We average $d\oplus l$ across all samples to get the LER of each circuit instance $\hat{\varepsilon}_L(c_j)$. Finally, we average over those to get the overall LER estimator $\hat{\varepsilon}_L$.}
\label{fig:ler_estimation}
\end{figure*}

\section{Derivation of the correctable volumes}
\label{ap:derivation_ft_regions}

\begin{figure*}[t]
    \centering
    \includegraphics{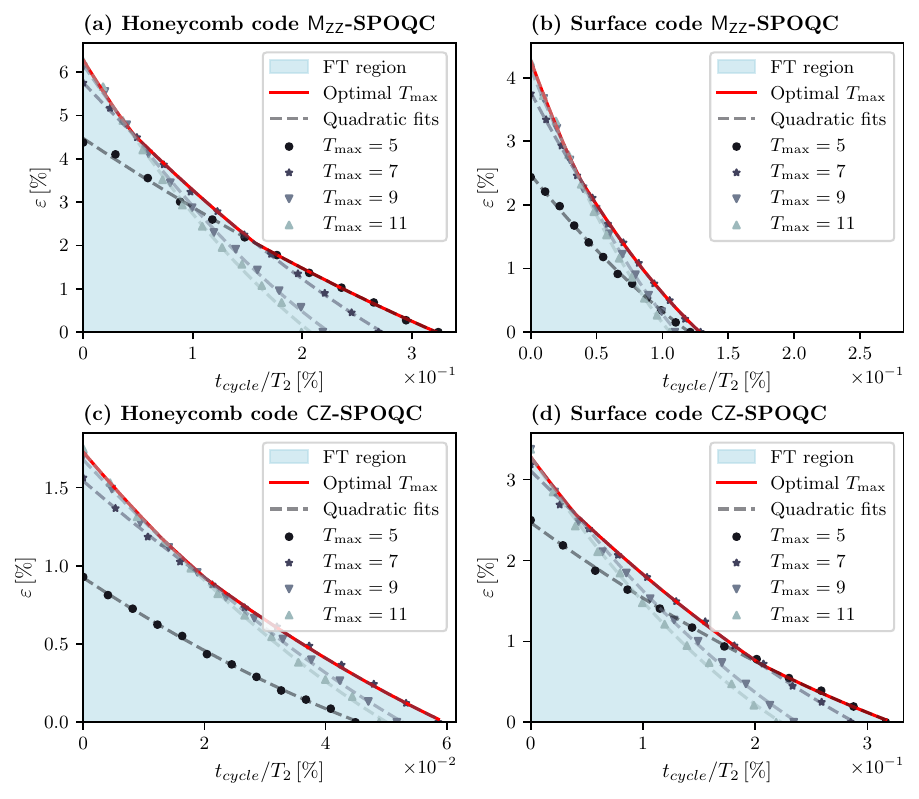}
    \caption{Fault-tolerant (FT) regions for pairs of codes and SPOQC implementations and $D=0$. The data-points represent simulated values of thresholds $x_{th}$ for different values of the maximum number of allowed trials $T_{\max}$. The dashed lines correspond to quadratic fits interpolating the data points and approximating $\Gamma_{0}$ for each value of $T_{\max}$. Finally, the red solid line corresponds to the fault-tolerant curve $\Gamma_0^{FT}$ obtained by taking the best possible threshold across values of $T_{\max}$.}
    \label{fig:all_ft_region}
\end{figure*}

This appendix is dedicated to the detailed derivation of the fault-tolerant regions and volumes.

\subsection{Fault-tolerant curves in the loss-decoherence plane}

We compare the different codes and architecture by exhibiting the \emph{fault-tolerant regions} in the loss-decoherence plane for a fixed value of the distinguishability $D$. In the following, we call $\tau=t_{cycle}/T_2$ the decoherence error parameter and take $D=0$, but the derivation is the same for other values of $D$. If we pick a point in parameter space $P^{(1)}=(\varepsilon^{(1)},\tau^{(1)}, 0)$ at unit distance from the origin, we can parametrized a straight line going from the origin to $P^{(1)}$ and past, with a single parameter $x^{(1)}\geq 0$. This parameter allows to have increasing noise along the direction defined by $P^{(1)}$ as $x^{(1)}\mapsto x^{(1)}P^{(1)}$. We can find a threshold $x^{(1)}_{th.}\geq 0$ on $x^{(1)}$. If we carry out the process with appropriately spread out points $(P^{(n)})_{1\leq n\leq N}$, we end up with a collection of $N$ threshold points $(x^{(n)}_{th.})_{1\leq n\leq N}$ from which we can extrapolate a curve $\Gamma_{0}$ in the loss, decoherence plane. Points $(\varepsilon_{th}, \tau_{th})$ on this curve correspond to $x_{th}(\varepsilon_{th}, \tau_{th})/\sqrt{\epsilon^2_{th}+\tau^2_{th}}$, i.e. threshold along the direction spanned by $(\varepsilon_{th}, \tau_{th})/\sqrt{\epsilon^2_{th}+\tau^2_{th}}$. 

For a given code, the curve described previously is also dependent on the maximum number of trials $T_{\max}$ allowed for the RUS gates. The points on the final fault-tolerant curve $\Gamma^{FT}_{D}$ are the best thresholds possible across $T_{\max}$. As an illustration, Fig.~\ref{fig:all_ft_region} shows the points $x_{th}^{(n)}$ and the extrapolated curves $\Gamma_0$ (dashed lines) and $\Gamma_0^{FT}$ (red lines), for $D=0$ and different pairs of codes and architecture. The fault-tolerant region is defined as the area below $\Gamma_0^{FT}$.

\subsection{Fault-tolerant surfaces and correctable volumes}

For a fixed value of $D$, we can associate a fault-tolerant curve $\Gamma_{D}^{FT}$ in the loss-decoherence plane. We define the fault-tolerant surface $\Sigma^{FT}$ in the loss, decoherence, distinguishability space through the equivalence
$$\Sigma^{FT}(\varepsilon,\tau,D)=0\Longleftrightarrow\Gamma_D^{FT}(\varepsilon,\tau)=0.$$
This is a two-dimensional surface inhabited by three-parameter thresholds. 
We can define a correctable volume $\mathrm{Vol}\left(\Sigma^{FT}\right)$ as the volume of the positive region below $\Sigma^{FT}$. For Fig.~\ref{fig:3d_ft_regions} of the main text, we computed the fault-tolerant surfaces by linear interpolation between the different displayed line fits using the \texttt{griddata} function of the \texttt{scipy} package. The volume of the Honeycomb Floquet code in Fig.~\ref{fig:3d_ft_regions}(a) is simply computed as a two-dimensional integral of the fault-tolerant surface. For the pentagon tilling in Fig.~\ref{fig:3d_ft_regions}(b), we proceed in several stages since we did not simulate past $D=0.075$ as the fault-tolerant regions were becoming too small. First, we compute the integral below the surface for $\varepsilon\in[\varepsilon_{th,u}, \varepsilon_{th,d}]$ (i.e. corresponding respectively to the loss threshold for $\tau=0$ and $D=0.075$ (highest green point) and the loss threshold for $\tau=0\%$ and $D=0\%$ (leftmost blue point)) and $\tau\in[\tau_{th,u}, \tau_{th,d}]$ (i.e. corresponding respectively to the decoherence threshold for $\varepsilon=0$ and $D=0.075$ (highest red point) and the decoherence threshold for $\varepsilon=0$ and $D=0$ (rightmost blue point)). We call this volume $\mathcal{V}_1$ and represent it as the schematic blue volume in Eq. \eqref{eq:volumes_pentagon}. Doing so, we are missing the volume $\mathcal{V}_2$ (red volume in Eq. \eqref{eq:volumes_pentagon}) of a trapezoid with two triangular bases $\mathcal{T}_1=\{(\varepsilon_{th,u},0,0), (0,0,0), (0,\tau_{th,u},0)\}$ and $\mathcal{T}_2=\{(\varepsilon_{th,u},0,0.075), (0,0,0.075), (0,\tau_{th,u},0.075)\}$. Finally, we approximate the volume of the remaining upper region $\mathcal{V}_3$ (green volume in Eq. \eqref{eq:volumes_pentagon}) by the volume of a pyramid with base $T_2$ and upper point $(0,0,1)$ (stemming from the insensitivity to the pentagon tiling to distinguishability errors alone). This approximation of $\mathcal{V}_3$ is optimistic since the fault-tolerant surface seems to display consistent convexity. Finally, the fault-tolerant volume for the pentagon tiling is approximately 
\begin{equation}
\begin{aligned}
\mathrm{Vol}\left(\Sigma^{FT}\right)&\approx\mathcal{V}_1+\mathcal{V}_2+\mathcal{V}_3\\
&=
\vcenter{\hbox{
\begin{tikzpicture}[scale=0.8]

  \coordinate (A) at (0,0,1.5);
  \coordinate (B) at (0,0,0.3);
  \coordinate (C) at (0,1,0.3);
  \coordinate (E) at (0.8,0,0);
  \coordinate (F) at (0.2,0,0);
  \coordinate (G) at (0.2,1,0);

  \fill[fill=blue!20,opacity=0.85] (G) -- (C) -- (B) -- (F) -- cycle;
  \filldraw[fill=blue!20,opacity=0.85, dashed] (A) -- (E) -- (F) -- (B) -- cycle; 
  \filldraw[fill=blue!20,opacity=0.85, dashed] (E) -- (F) -- (G) -- cycle; 
  \filldraw[fill=blue!20,opacity=0.85, dashed] (A) -- (B) -- (C) -- cycle; 
  \filldraw[fill=blue!20,opacity=0.85] (A) -- (E) -- (G) -- (C) -- cycle;  
\end{tikzpicture}}}+\vcenter{\hbox{
\begin{tikzpicture}[scale=0.8]

  \coordinate (A) at (0,0,0);
  \coordinate (B) at (0,0,0.3);
  \coordinate (C) at (0,1,0.3);
  \coordinate (E) at (0,1,0);
  \coordinate (F) at (0.2,0,0);
  \coordinate (G) at (0.2,1,0);

  \draw[dashed,opacity=0.85] (A) -- (E);
  \draw[dashed,opacity=0.85] (A) -- (B);
  \draw[dashed,opacity=0.85] (A) -- (F);
  \filldraw[fill=red!20,opacity=0.85, dashed] (A) -- (B) -- (F) -- cycle;
  \filldraw[fill=red!20,opacity=0.85] (B) -- (C) -- (G) -- (F) -- cycle;
  
  \filldraw[fill=red!20,opacity=0.85] (E) -- (G) -- (C) -- cycle;
\end{tikzpicture}}}+
\vcenter{\hbox{
\begin{tikzpicture}[scale=0.8]

  \coordinate (A) at (0,1.1,0);
  \coordinate (C) at (0,0,0.3);
  \coordinate (E) at (0,0,0);
  \coordinate (G) at (0.2,0,0);

  \draw[dashed,opacity=0.85, dashed] (E) -- (C);
  \draw[dashed,opacity=0.85, dashed] (E) -- (G);
  \draw[dashed,opacity=0.85, dashed] (E) -- (A);
  \fill[fill=green!20,opacity=0.5] (E) -- (C) -- (G) -- cycle;
  \fill[fill=green!20,opacity=0.5] (E) -- (C) -- (A) -- cycle;
  \fill[fill=green!20,opacity=0.5] (E) -- (G) -- (A) -- cycle;
  \fill[fill=green!20,opacity=0.5] (A) -- (G) -- (C) -- cycle;
  \draw[opacity=0.85] (A) -- (G) -- (C) -- cycle;
\end{tikzpicture}}}\\
&=\vcenter{\hbox{
\begin{tikzpicture}[scale=0.8]

  \coordinate (A) at (0,2.1,0);
  \coordinate (C) at (0,1,0.3);
  \coordinate (E) at (0,1,0);
  \coordinate (G) at (0.2,1,0);

  \draw[dashed,opacity=0.85, dashed] (E) -- (C);
  \draw[dashed,opacity=0.85, dashed] (E) -- (G);
  \draw[dashed,opacity=0.85, dashed] (E) -- (A);
  \fill[fill=green!20,opacity=0.5] (E) -- (C) -- (G) -- cycle;
  \fill[fill=green!20,opacity=0.5] (E) -- (C) -- (A) -- cycle;
  \fill[fill=green!20,opacity=0.5] (E) -- (G) -- (A) -- cycle;
  \fill[fill=green!20,opacity=0.5] (A) -- (G) -- (C) -- cycle;
  \draw[opacity=0.85] (A) -- (G) -- (C) -- cycle;

  \coordinate (A) at (0,0,0);
  \coordinate (B) at (0,0,0.3);
  \coordinate (C) at (0,1,0.3);
  \coordinate (E) at (0,1,0);
  \coordinate (F) at (0.2,0,0);
  \coordinate (G) at (0.2,1,0);

  \draw[dashed,opacity=0.85] (A) -- (E);
  \draw[dashed,opacity=0.85] (A) -- (B);
  \draw[dashed,opacity=0.85] (A) -- (F);
  \filldraw[fill=red!30,opacity=0.85, dashed] (A) -- (B) -- (F) -- cycle;
  \filldraw[fill=red!30,opacity=0.85] (B) -- (C) -- (G) -- (F) -- cycle;
  
  \filldraw[fill=red!30,opacity=0.85] (E) -- (G) -- (C) -- cycle;

  \coordinate (A) at (0,0,1.5);
  \coordinate (B) at (0,0,0.3);
  \coordinate (C) at (0,1,0.3);
  \coordinate (E) at (0.8,0,0);
  \coordinate (F) at (0.2,0,0);
  \coordinate (G) at (0.2,1,0);

  \fill[fill=blue!20,opacity=0.85] (G) -- (C) -- (B) -- (F) -- cycle;
  \filldraw[fill=blue!20,opacity=0.85, dashed] (A) -- (E) -- (F) -- (B) -- cycle; 
  \filldraw[fill=blue!20,opacity=0.85, dashed] (E) -- (F) -- (G) -- cycle; 
  \filldraw[fill=blue!20,opacity=0.85, dashed] (A) -- (B) -- (C) -- cycle; 
  \filldraw[fill=blue!20,opacity=0.85] (A) -- (E) -- (G) -- (C) -- cycle;  
\end{tikzpicture}}},
\end{aligned}
\tag{C1}
\label{eq:volumes_pentagon}
\end{equation}
where in the last equality, the red volume can be found behind the blue one. The blue and red shapes sketch out the volume displayed in Fig.~\ref{fig:3d_ft_regions}(a) in the main text while the green one approximate the remaining upper volume not displayed in Fig.~\ref{fig:3d_ft_regions}(a).

\section{Additional threshold plots}
\label{ap:other_thr_plots}

\begin{figure*}
    \centering
    \includegraphics{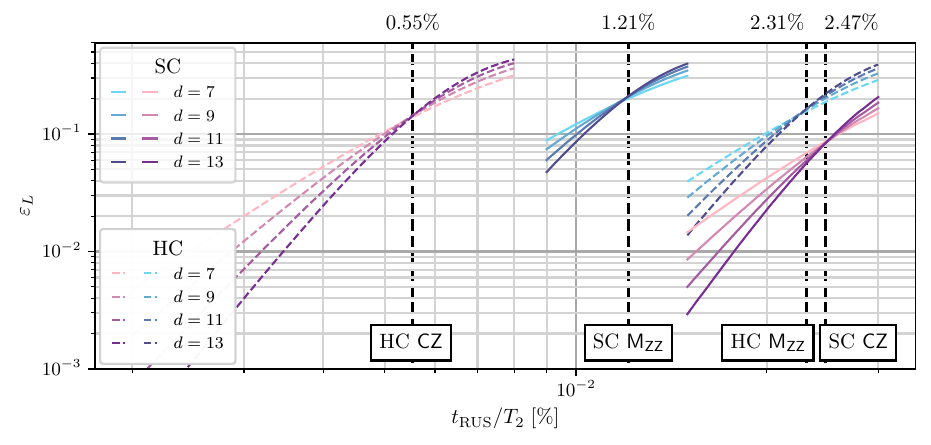}
    \caption{Logical error rate $\varepsilon_L$ as a function of decoherence time $t_{\rm RUS}/T_2$ for different distances in the $\CZ$-SPOQC architecture (pink gradient) and $\MZZ$-SPOQC architecture (blue gradient). Thresholds values are reported on the top of the plots and marked by the black dashed lines. These curves are obtained by averaging out the mismatch between the measure of the logical observable $\X$ obtained through a data readout at the end of the QEC cycles, and the predicted observable by the MWPM decoder \cite{higgott2023sparse}.}
    \label{fig:deco_thresholds}
\end{figure*}

\begin{figure*}
    \centering
    \includegraphics{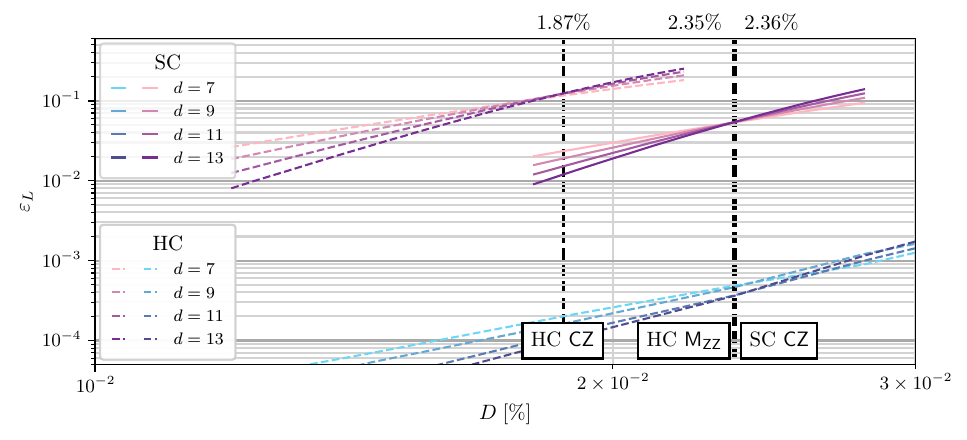}
    \caption{Logical error rate $\varepsilon_L$ as a function of photon distinguishability $D$ for different distances in the $\CZ$-SPOQC architecture (pink gradient) and $\MZZ$-SPOQC architecture (blue gradient). Thresholds values are reported on the top of the plots and marked by the black dashed lines. These curves are obtained by averaging out the mismatch between the measure of the logical observable $\X$ obtained through a data readout at the end of the QEC cycles, and the predicted observable by the MWPM decoder \cite{higgott2023sparse} (for both the honeycomb code and surface code in $\CZ$-SPOQC architectures), and the BP+OSD decoder \cite{roffe_decoding_2020} (for the honeycomb code in the $\MZZ$-SPOQC architecture).}
    \label{fig:dist_thresholds}
\end{figure*}

In this section, Fig. \ref{fig:deco_thresholds} and \ref{fig:dist_thresholds} show the logical error rates plot for decoherence and distinguishability induced errors discussed in Sections \ref{sec:deco_thr} and \ref{sec:dist_thr} in the main text.

\bibliographystyle{quantum}
\bibliography{bibliography}

\end{document}